\documentclass[a4paper,12pt]{article}

\usepackage{amsmath}
\usepackage{amssymb}
\usepackage{color}
\usepackage{graphics}
\usepackage{graphicx,epsfig}
\usepackage{amsfonts}

\usepackage{inputenc}
\inputencoding{latin1}

  \newcommand{\bsg}{BR(b\rightarrow s \gamma)}
  \newcommand{\DM}{\Omega_{CDM}h^2}
  \newcommand{\DMW}{\Omega^{WMAP}_{CDM}h^2}
  \newcommand{\gmu}{\delta a_{\mu}}
  \newcommand{\nn}{\nonumber}
  \newcommand{\stau}{\tilde{\tau}}
  \newcommand{\sel}{\tilde{e}_R}
  \newcommand{\smu}{\tilde{\mu}_R}
  \newcommand{\neut}{\tilde{\chi}^0_1}

  \newcommand{\sneut}{\tilde{\nu}}

  \newcommand{\DeltaO}{\Delta^{\Omega}}
  \newcommand{\DeltaEW}{\Delta^{EW}}
  \newcommand{\tanb}{\tan\beta}
  \newcommand{\sigI}{1\sigma}
  \newcommand{\sigII}{2\sigma}

\begin{document}

\begin{titlepage}

\begin{flushright}
CERN-PH-TH/2007-211 \\
November 2007
\end{flushright}

  \begin{center}
    {
      {
    \bigskip{} {\par\centering \textbf{\large The Fine-Tuning
        Price of Neutralino Dark Matter in Models with Non-Universal
        Higgs Masses} \large \par}
    \bigskip{}

        {\par\centering John Ellis$^{1}$, S. F. King$^{2}$ and
          J. P. Roberts$^{3}$\\ \par}
        \bigskip{}

            {\par\centering
              {\small $^1$ Theory Division, Physics department, CERN}\\
              {\small $^2$ School of Physics and Astronomy,
                University of Southampton}\\
              {\small $^3$ Institute of Theoretical Physics, Warsaw University}\\
              \par}

            \bigskip{}

            \begin{abstract}
              \noindent
We study the amounts of fine-tuning of the parameters of the
MSSM with non-universal soft supersymmetry-breaking contributions
to the Higgs masses (the NUHM) that would be required for the
relic neutralino density to lie within the range favoured by WMAP
and other astrophysical and cosmological observations.
Such dark matter fine-tuning is analogous to the commonly
studied electroweak fine-tuning associated
with satisfying the electroweak symmetry breaking conditions,
which we also study for completeness.
We identify several distinct regions of the NUHM parameter space: a bulk
region, a $\stau-\neut$ coannihilation region, a pseudoscalar
Higgs funnel region, a focus-point bino/higgsino region and a
$\sneut-\neut$ coannihilation region. Within each region, we
analyse specific representative points for which we provide
breakdowns of the contributions to the dark matter fine-tuning
associated with the different NUHM parameters. In general, the
NUHM offers points with both smaller and larger amounts of
dark matter fine-tuning than points in the corresponding regions of the CMSSM.
Lower amounts of dark matter fine-tuning typically arise at points where
several different (co)annihilation processes contribute, e.g., at
junctions between regions with different dominant processes. We
comment on the prospects for using collider measurements to
estimate the likely dark matter density within the NUHM framework.
            \end{abstract}
      }
    }
  \end{center}
\end{titlepage}

\newpage

\setcounter{footnote}{0}

\section{Introduction}
\label{Intro}

The primary utilitarian motivation for supersymmetry being accessible
to experiments at the electroweak scale, e.g., at the LHC, depends on
its ability to alleviate the problem of fine-tuning of electroweak
symmetry breaking present in the Standard Model
\cite{EENZ,hep-ph/0312378}.  A supplementary phenomenological
motivation for weak-scale supersymmetry is its ability to provide the
cold dark matter required by astrophysics and cosmology \cite{EHNOS,
hep-ph/9506380}. This is a natural feature of supersymmetric models
that conserve R parity, with the lightest neutralino $\neut$ being
particularly well-suited to provide the preferred amount of cold dark
matter if it is the lightest supersymmetric particle (LSP) and weighs
less than about 1~TeV \cite{EHNOS,etcEllis:1985jn}. Within the general
supersymmetric framework, one may find more plausible regions of the
supersymmetric parameter space that are less fine-tuned, in the sense
that the values of the model parameters chosen at some high input
scale require less delicate adjustment in order to obtain the correct
value of the electroweak scale \cite{EENZ,Barbieri:1987fn}, as
measured by $M_Z$, or the correct value of the cold dark matter
density $\Omega_{CDM}h^2$
\cite{Ellis:2001zk,hep-ph/0603095,hep-ph/0608135,King:2007vh}.

It is hard to make this type of plausibility argument at all rigorous:
it is notoriously difficult to make probabilistic statements about the
unique (by definition) Universe in which we live, it is largely a
matter of personal choice which derived quantity one should consider
and which input parameters one wishes to avoid fine-tuning, it is
difficult to argue conclusively for the superiority of one measure of
fine-tuning over any other, and even less easy to agree on a `pain
threshold' in the amount of fine-tuning one is prepared to tolerate
\cite{Barbieri:1987fn}. Nevertheless, within a given model framework
with its specific input parameters, it is legitimate to consider some
important derived quantity such as $\Omega_{CDM}h^2$, and compare the
amounts of fine-tuning required in different regions of its parameter
space, which frequently do not depend very sensitively on the specific
sensitivity measure employed.

Moreover, even if one does not accept that the less sensitive
parameter regions are more plausible, measuring the dark matter
fine-tuning may have other uses. For example, one hopes (expects) some
day to discover supersymmetry and start to measure the values of its
parameters. Unavoidably, these will have non-negligible measurement
errors, and these uncertainties propagate via the dark matter
fine-tuning parameters into the calculation, e.g., of
$\Omega_{CDM}h^2$. One of the key features of supersymmetry is its
ability to provide a calculable amount of cold dark matter, and it is
interesting to know how accurately which of its parameters must be
measured in order to calculate $\Omega_{CDM}h^2$ with an accuracy
comparable to that quoted by astrophysicists and cosmologists
\cite{Ellis:2001zk,Battaglia:2003ab}. An accurate calculation of
$\Omega_{CDM}h^2$ might also reveal some deficiency of the
supersymmetric explanation of the cold dark matter, and possibly the
need for some other new physics in addition.

It is important to note that the parameters we refer to here are the
GUT-scale soft supersymmetry-breaking masses and couplings, whereas
experiments would measure directly physical masses and mixings at much
lower energies. Ideally, one would calculate the relic density
directly from the low-energy measurements of MSSM parameters. However
it will be difficult, if not impossible, to pin down all the key
parameters using LHC data alone, except by making supplementary
assumptions about the pattern of supersymmetry breaking at the GUT
scale, as we do here. Assuming a structure of GUT-scale unification,
one may use experimental measurements to constrain these fewer
high-energy parameters. The strength of the constraints will depend on
the magnitudes of these parameters and the experimental tools
available. Very likely some accelerator beyond the LHC will be needed,
but we do not yet know what will be available.  The fine-tuning
measures we calculate here show clearly which of the high-energy
parameters are most important for a precise calculation of the relic
density, and hence contribute to the `wish list' for such an
accelerator.

For these reasons, we make no further apologies for considering the
fine-tuning of $\Omega_{CDM}h^2$ in this paper, which we shall refer
to as ``dark matter fine-tuning''
\cite{Ellis:2001zk,hep-ph/0603095,hep-ph/0608135,King:2007vh} to
distinguish it from the more commonly studied ``electroweak
fine-tuning'' \cite{EENZ,Barbieri:1987fn}, which we also consider for
completeness. The issue of dark matter fine-tuning has been considered
previously in the context of several different models including the
constrained minimal supersymmetric extension of the Standard model
(CMSSM) \cite{Ellis:2001zk}, in which the soft supersymmetry-breaking
scalar masses $m_0$, gaugino masses $m_{1/2}$ and trilinear parameters
$A_0$ are each assumed to be universal, a more general MSSM with
non-universal third family scalars and gaugino masses
\cite{hep-ph/0603095}, a string-inspired non-universal model
\cite{hep-ph/0608135} and SUSY GUTs with non-universal gaugino masses
\cite{King:2007vh}. Here we extend such considerations to models with
non-universal soft supersymmetry-breaking contributions to the Higgs
masses (NUHM).  Within this NUHM framework, the independent input
parameters may be taken as \cite{oldnuhm,Ellis:2002wv,hep-ph/0210205}
\begin{equation}
  a_{NUHM}=\left\{m_0,~m_{H_1},~m_{H_2},~m_{1/2},~A_0,~\tanb,
  ~\text{sign}(\mu)\right\},
\end{equation}
and we take as our measure of dark matter fine-tuning the quantity
\begin{equation}
  \Delta_\Omega \; \equiv \; {\rm Max}_i \left| \frac{a_i}{\Omega_\chi}
  \frac{\partial \Omega_\chi}{\partial a_i} \right| .
\end{equation}
Our objective will be three-fold: to compare the amount of dark matter
fine-tuning required within the NUHM to that required within the
CMSSM, to identify the regions of the NUHM parameter space that
require relatively less (or more) dark matter fine-tuning, and thereby
to quantify the accuracy in the determination of the GUT-scale NUHM
parameters that would be needed in order to calculate $\Omega_\chi
h^2$ with any desired accuracy.

The regions of the NUHM parameter space where $\Omega_\chi h^2$ falls
within the range favoured by WMAP and other experiments has been
studied quite extensively, for example in \cite{hep-ph/0210205}.  It
shares several features in common with the more restrictive CMSSM
framework proposed in \cite{Kane:1993td} and extensively studied in
\cite{Ellis:1999mm}.  For example, there are regions where $\neut$ -
stau coannihilation is important, and others where $\neut$ pairs
annihilate rapidly via direct-channel $H, A$ poles. However, other
possibilities also occur. For example, there are regions where $\neut$
- sneutrino coannihilation is dominant. Also there are regions where
rapid-annihilation and bulk regions, which are normally separated by a
coannihilation strip, approach each other and may even merge. As we
discuss below in more detail, the sneutrino coannihilation regions
exhibit relatively high dark matter fine-tuning, whereas the `merger'
regions may require significantly less dark matter fine-tuning.

In this work we provide a first calculation of the dark matter
fine-tuning for the regions of the NUHM that are favoured by dark
matter measurements. In addition, we present a first calculation of
the electroweak fine-tuning within this model and update the parameter
scans for the current measurement of the top mass.

The rest of the paper is laid out as follows. In
Section~\ref{sec:methods} we summarise the methods used in our
numerical studies. Next, in Section~\ref{CMSSM} we review the familiar
case of the CMSSM, which serves as a baseline for later
comparison. Then, in Section~\ref{NUHM} we study dark matter within
the NUHM model in which universality between the soft
supersymmetry-breaking masses of the sfermions (squarks and sleptons)
and Higgs multiplets is broken. Finally, in Section~\ref{Conc} we
present our conclusions.

\section{Methodology}
\label{sec:methods}

\subsection{Codes}

In order to study the low-energy phenomenology of the NUHM, we need a
tool to run the mass spectrum from the GUT scale down to the
electroweak scale using the renormalisation group equations
(RGEs)\cite{Martin:1993zk}. For this purpose we use the RGE code {\tt
SoftSusy}~\cite{hep-ph/0104145}. This interfaces with the MSSM package
within {\tt micrOMEGAs}~\cite{hep-ph/0112278}, which we use to
calculate the dark matter relic density $\DM$, $\bsg$ and $\gmu$. We
take $m_t=170.9$~GeV throughout.

\subsection{Theoretical, Experimental and Cosmological Bounds}

After running the mass spectrum of any chosen model parameter set from
the GUT scale down to the electroweak scale, we perform a number of
checks on the phenomenological acceptability of the point chosen. A
point is ruled out if:
\begin{enumerate}
\item It does not provide radiative electroweak symmetry breaking
  (REWSB). Such regions are displayed in light red in the subsequent
  figures.
\item It violates the bounds on particle masses provided by the
  Tevatron and LEP~2. Such regions are displayed in light
  blue~\footnote{The current LEP~2 bound on the lightest MSSM Higgs
  stands at $114.4$~GeV. However, there is a theoretical uncertainty
  of some $3$~GeV in the determination of the mass of the light
  Higgs~\cite{Allanach:2004rh}. Rather than placing a hard cut on the
  parameter space for the Higgs mass, instead we plot a line at
  $m_h=111$~GeV and colour the region in which $m_h <111$~GeV in very
  light grey-blue.}.
\item It results in a lightest supersymmetric particle (LSP) that is
  not the lightest neutralino. We colour these regions light green.
\end{enumerate}

In the remaining parameter space we display the 1- and 2-$\sigma$
regions for $\gmu$ and $\bsg$, as well as plotting the 2-$\sigma$
region for the relic density allowed by WMAP and other observations.

\subsubsection{$\gmu$}

Present measurements of the anomalous magnetic moment of the muon
$a_\mu$ deviate from theoretical calculations of the SM contribution
based on low-energy $e^+ e^-$ data~\footnote{There is a long-running
debate whether the calculation of the hadronic vacuum polarisation in
the Standard Model should be done with $e^+e^-$ data, or with $\tau$
decay data. The weight of evidence indicates the $e^+e^-$ estimate is
more reliable so we use the $e^+ e^-$ value in our work.}. Taking the
current experimental world average and the state-of-the-art SM value
from~\cite{hep-ph/0703049}, there is a discrepancy:
\begin{equation}
  (a_\mu)_{exp}-(a_\mu)_{SM}=\delta a_\mu = (2.95\pm
  0.88)\times 10^{-9},
\end{equation}
which amounts to a 3.4-$\sigma$ deviation from the SM value.  As
already mentioned, we use {\tt micrOMEGAs} to calculate the SUSY
contribution to $(g-2)_\mu$. The dominant theoretical errors in this
calculation are in the SM contribution, so in our analysis we neglect
the theoretical error in the calculation of the SUSY contribution.

\subsubsection{$\bsg$}

The variation of $\bsg$ from the value predicted by the Standard Model
is highly sensitive to SUSY contributions arising from charged
Higgs-top loops and chargino-stop loops. To date no deviation from the
Standard Model has been detected. We take the current world average
from~\cite{hfag}, based on the BELLE~\cite{hep-ex/0103042},
CLEO~\cite{hep-ex/0108033} and BaBar~\cite{Aubert:2005cu}
measurements:
\begin{equation}
  \bsg = (3.55 \pm 0.26) \times 10^{-4}.
\end{equation}
Again, we use {\tt micrOMEGAs} to calculate both the SM value of
$\bsg$ and the SUSY contributions. It is hard to estimate the
theoretical uncertainty in the calculation of the SUSY contributions,
but note that there is an uncertainty of $10\%$ in the NLO SM
prediction of $\bsg$~\cite{bsgNLO}~\footnote{We recall that {\tt
micrOMEGAs} calculates the SM contribution to $\bsg$ to NLO. A first
estimate of the SM prediction of $\bsg$ to NNLO was presented
in~\cite{Misiak:2006zs}. This showed a decrease of around $0.4\times
10^{-4}$ in the central value of the SM prediction. The implementation
of the NNLO contributions in the calculation is non-trivial and its
implementation in {\tt micrOMEGAs} is currently underway. As a result
we do not include this decrease in the results we present, but instead
note that \textit{positive} SUSY contributions to $\bsg$ look likely
to be favoured in future. This would favour a negative sign of $\mu$
and thus cause tension with $(g-2)_\mu$.}. As with $\gmu$, we plot the
1-$\sigma$ and 2-$\sigma$ experimental ranges, and do not include a
theoretical error in the calculation.

\subsubsection{$\DM$}
Evidence from the cosmic microwave background, the rotation curves of
galaxies and other astrophysical data point to a large amount of cold
non-baryonic dark matter in the universe. The present
measurements~\cite{astro-ph/0603449} indicate the following value for
the current cold dark matter density:
\begin{equation}
  \DM = 0.106 \pm 0.008.
\end{equation}
We calculate the relic dark matter density with {\tt micrOMEGAs} using
the \textit{fast} approximation. Given a low-energy mass spectrum,
{\tt micrOMEGAs} gives an estimated precision of $1\%$ in the
theoretical prediction of the relic density. This is negligible
compared to the present observational error, so the 2-$\sigma$ band
plotted takes into account only the experimental error~\footnote{We
emphasize that the quoted $1\%$ accuracy is for a given low-energy
spectrum, which is obtained using {\tt softsusy}.  However, there are
differences in the details of the mass spectrum between
codes~\cite{Allanach:2003jw}, for given high-energy inputs, and
different dark matter regions have different levels of sensitivity to
these variations: see~\cite{Belanger:2005jk} for a detailed study. The
result of the discrepancies between codes is to move the dark matter
regions slightly in the GUT scale parameter space. As we are
interested in broad features of these regions, rather than their
precise locations, these uncertainties are not important for our
purposes.}.

In the following Sections, we calculate the dark-matter fine-tuning
for any point that lies within the $\sigII$ allowed region, and
indicate the amount using colour coding. We also display electroweak
fine-tuning contours over the different regions.

\section{The Constrained Minimal Supersymmetric Standard Model}
\label{CMSSM}

We first review the familiar Constrained Minimal Supersymmetric
Standard Model (CMSSM) \cite{Kane:1993td,Ellis:1999mm}. 
which serves as a standard to which we compare
the parameter space of the NUHM.

The CMSSM has a much simpler spectrum of soft masses than the full
MSSM. First, all of the soft squark and slepton (mass)$^2$ matrices
are chosen to be diagonal and universal at the GUT scale with the
diagonal entries equal to $m_0^2$. Secondly both the soft Higgs
(mass)$^2$ are also set equal to $m_0^2$. Additionally, all the
gaugino masses are assumed to be unified with a value $m_{1/2}$ at the
GUT scale. Finally, we take the trilinear coupling matrices to have
only one non-zero entry (the third-family dominance approximation) and
assume that all these entries are equal to a common value
$A_0$. Requiring that electroweak symmetry be broken radiatively to
give the observed electroweak boson masses, we trade the soft
parameters $\mu$ and $B$ for $\tanb$, the ratio of the Higgs vevs, and
the sign of $\mu$. This results in a model with four free parameters
and a sign:
\begin{equation}
  a_{CMSSM} \in \left\{
  m_0,~m_{1/2},~A_0,~\tanb,~\text{sign}(\mu)\right\}.
\end{equation}
Although our main focus is the dark-matter fine-tuning, we also report
the required amounts of electroweak fine-tuning for
specific cases of interest.

\begin{figure}[ht!]
  \begin{center}
    \scalebox{1.0}{\includegraphics{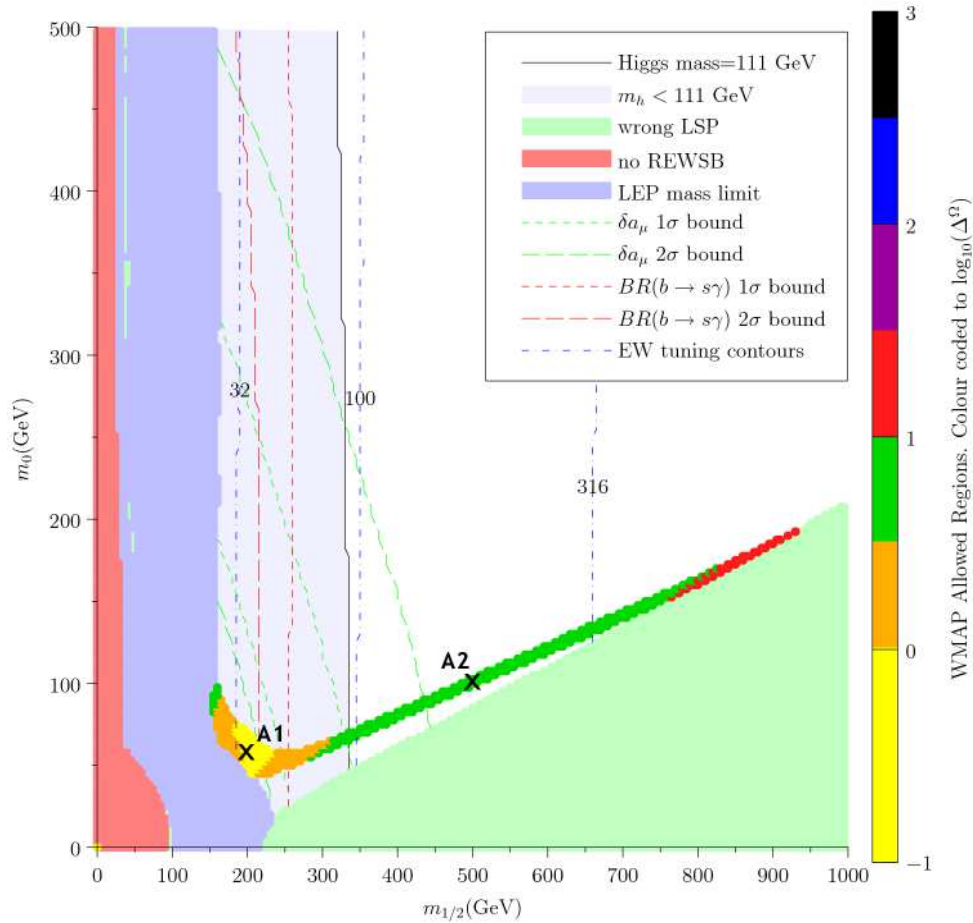}}
  \end{center}
  \vskip -0.5cm \caption{\small The $(m_{1/2},m_0)$ plane of the CMSSM
     with $A_0=0,~\tanb=10$ and sign$(\mu)$
    +ve.\label{f:CMSSM,t10}}
\end{figure}

\begin{figure}[ht!]
  \begin{center}
    \scalebox{1.0}{\includegraphics{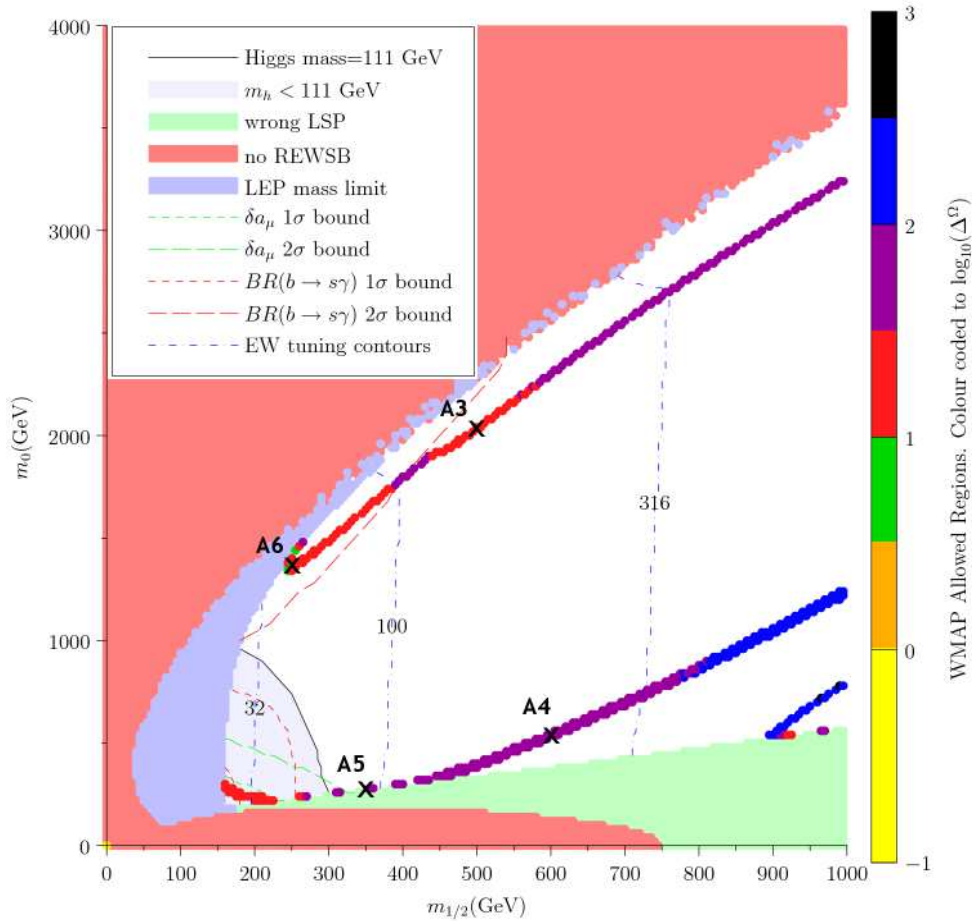}}
  \end{center}
  \vskip -0.5cm \caption{\small The $(m_0,~m_{1/2})$ plane of the
    CMSSM parameter space for $A_0=0$, $\tanb=50$ and sign($\mu$)
    +ve.\label{f:CMSSM}}
\end{figure}

In Fig.~\ref{f:CMSSM,t10} we show the $(m_0,m_{1/2})$ plane of the
CMSSM for $A_0=0,~\tanb=10$ and sign$(\mu)$ positive. At low $m_0$ the
parameter space is ruled out because $m_{\stau}<m_{\neut}$, resulting
in a stau LSP (light green). Regions at low $m_{1/2}$ are ruled out by
LEP~2 bounds on the masses of the charginos and sleptons (light
blue). A larger range of $m_0$ is ruled out by the absence of a light
Higgs boson, and we shade the region with $m_h<111$~GeV (light grey
with a black boundary). Finally, the model fits the current $\gmu$
measurement for low $m_0$ and $m_{1/2}$. The $\sigI$ and $\sigII$
bounds are shown as a short and long-dashed green lines
respectively. At larger $m_0$ and $m_{1/2}$, the SUSY contribution
becomes small as the sparticles that contribute in loops become
heavy. Therefore $\gmu\neq 0$ favours relatively light soft masses.

In the remainder of the parameter space we find the regions that fit
the WMAP strip at $\sigII$, and for each such point we calculate the
fine-tuning of the dark matter density. Each point is then plotted
with a colour that corresponds to the value of the tuning via the
log-scale on the right hand side.

\begin{table}
  \begin{center}
    \begin{tabular}{|l|l|l|l|l|}
      \hline
      \multicolumn{1}{|c|}{Parameter} &
      \multicolumn{2}{|c|}{A1} &
      \multicolumn{2}{|c|}{A2}\\
      \cline{2-5}
                & value & $\DeltaO$ & value & $\DeltaO$ \\
      \hline
      $m_0$     & 60    & 0.62 & 100 & 5.7\\
      $m_{1/2}$ & 200   & 0.99 & 500 & 5.8\\
      $\tanb$   & 10    & 0.13 & 10  & 1.5\\
      \hline
      $\Delta_\Omega$ & & 0.99 &     & 5.8\\
      \hline
      $\Delta_{EW}$   & & 37   &     & 190\\
      \hline
    \end{tabular}
  \end{center}\vskip -0.5cm
  \caption{\small A summary of the properties of points A1 and A2,
    shown in Fig.~\ref{f:CMSSM,t10}, chosen as representatives of the
    bulk region (A1) and stau-coannihilation region (A2). We also
    present a breakdown of the dark-matter and electroweak
    fine-tunings with respect to each parameter of the CMSSM, except
    for $A_0$, which we fix: $A_0=0$ here and elsewhere in this
    paper.\label{t:CMSSM,t10}}
\end{table}

There are two distinct regions that fit the WMAP measurement of the
relic density. The first region lies at $m_{1/2}\approx 200$~GeV,
$m_0\approx 70$~GeV and contains the point A1. This is the bulk
region, which is adjacent to the light blue region that is ruled out
by the LEP~2 slepton mass constraint. The lightness of the sleptons
enhances neutralino annihilation via $t$-channel slepton exchange to
the extent that it allows bino dark matter to fit the WMAP relic
density measurement. This process is relatively insensitive to the
precise masses of the neutralino and the sleptons. This is reflected
in the fact that much of the bulk region is plotted in yellow,
signifying $\DeltaO<1$. From the breakdown of the tunings of point A1
in Table~\ref{t:CMSSM,t10} it is clear that the tuning is mainly in
the parameters $m_{1/2}$ and $m_0$. The sensitivity to $m_0$ is to be
expected, as $t$-channel slepton exchange is clearly dependent on the
mass of the exchanged particle. This also explains the sensitivity to
$m_{1/2}$ since, although the masses of the sleptons are determined by
$m_0$ at the GUT scale, the running to low energies is dominated by
$m_{1/2}$. Thus, in this region the masses of the light sleptons are
sensitive to both $m_0$ and $m_{1/2}$. The relic density is also
dependent upon the mass of the LSP, in this case (mainly) a bino with
a mass determined primarily by $m_{1/2}$. It is therefore apparent why
the sensitivity of the bulk regions lies primarily with $m_0$ and
$m_{1/2}$. The fact that no precise cancellations or balancing of
parameters is required explains why the tuning is low. However, the
bulk region lies inside a light grey region signifying a Higgs with a
mass less than $111$~GeV. Therefore we do not consider this region
further here.

The second region that fits the observed relic density is the band
that contains point A2. This band lies alongside the area that is
ruled out by a stau LSP (light green). Along the edge of the light
green region, the mass of the stau is close to that of the lightest
neutralino, resulting in comparable number densities of both particles
around the time of freeze-out. Thus many more annihilation channels
must be considered, such as the annihilations of $\stau-\stau$ and
$\stau-\neut$ in addition to $\neut-\neut$. This suppresses the number
density of SUSY particles and thus reduces the resultant dark matter
relic density. The effect of coannihilation depends strongly upon the
number density of NLSPs at freeze-out. This is in turn very sensitive
to the mass difference between the NLSP and the LSP. Thus, it may at
first sight seem surprising that the $\stau$ coannihilation band is
plotted here in green and red, corresponding to relatively low dark
matter fine-tuning $\DeltaO=3-11$. The reason, as discussed
in~\cite{hep-ph/0603095}, is that in this region the RGE running
results in the mass of the right handed $\stau$ and the mass of the
lightest neutralino both being dominantly dependent on $m_{1/2}$, with
the stau having a smaller secondary dependence on $m_0$. As a result,
their masses vary together as the soft masses are varied, and $\Delta
m$ is remarkably insensitive to $m_0$ or $m_{1/2}$. This mitigates the
normal sensitivity of the coannihilation region. We take the point A2
as a representative point in this band and provide a breakdown of the
individual tunings in Table~\ref{t:CMSSM,t10}. The tuning is equally
dependent upon $m_0$ and $m_{1/2}$, as expected.

As well as the dark-matter fine-tuning, for reference we also
calculate the fine-tuning required to fit the electroweak boson
masses. This calculation is performed across the parameter space using
the same measure as we use for the dark matter fine-tuning. We plot
the tuning using blue dot-dashed contours in Fig.~\ref{f:CMSSM,t10},
and label each contour with their respective values of
$\DeltaEW=\text{max}(\DeltaEW_a)$. We also list the value of
$\DeltaEW$ in the last row of Table~\ref{t:CMSSM,t10} for both points
we consider.  The $W$ boson mass is in general the result of a careful
balancing act between the soft masses: the larger the soft masses, the
more precisely they must cancel to give the $W$ boson mass. Therefore
it is unsurprising that the amount of fine-tuning required to obtain
the correct electroweak symmetry breaking increases smoothly with the
increasing soft masses. The electroweak and dark matter fine-tunings
are largely independent.

Other regions in which the CMSSM fits the observed relic density are
seen when we consider larger values of $\tanb$. In Fig.~\ref{f:CMSSM}
we consider a wider range in the $(m_0,m_{1/2})$ plane and take
$\tanb=50$.

Low $m_0$ is once again ruled out because the stau is the LSP (light
green). Equally low $m_{1/2}$ and $m_0$ are ruled out by searches for
the lightest Higgs (light grey with a black boundary) or LEP~2 bounds
on the chargino and light sleptons (solid light blue). Finally large
$m_0$ and low $m_{1/2}$ are ruled out as they do not give REWSB (light
red).  We plot the different regions that fit the relic density using
the same colour coding for dark-matter fine-tuning as was used in
Fig.~\ref{f:CMSSM,t10}.

\begin{table}
  \begin{center}
    \begin{tabular}{|l|l|l|l|l|l|l|l|l|}
      \hline
      \multicolumn{1}{|c|}{Parameter} &
      \multicolumn{2}{|c|}{A3} &
      \multicolumn{2}{|c|}{A4} &
      \multicolumn{2}{|c|}{A5} &
      \multicolumn{2}{|c|}{A6}\\
      \cline{2-9}
                & value & $\DeltaO$ & value & $\DeltaO$ & value & $\DeltaO$ & value
& $\DeltaO$ \\
      \hline
      $m_0$     & 2030  & 27   & 540 & 5.0  & 277   & 19   & 1400& 13   \\
      $m_{1/2}$ & 500   & 18   & 600 & 8.0  & 350   & 12   & 250 & 22   \\
      $\tanb$   & 50    & 2.0  & 50  & 76   & 50    & 48   & 50  & 8.2  \\
      \hline
      $\Delta_\Omega$ & & 27   &     & 76   &       & 48   &     & 22   \\
      \hline
      $\Delta_{EW}$  &  & 150  &     & 230  &       & 92   &     & 48   \\
      \hline
    \end{tabular}
  \end{center}\vskip -0.5cm
  \caption{\small Here we summarize properties of the points A3-6,
    shown in Fig.~\ref{f:CMSSM}, which represent the higgsino/bino
    region (A3,6), the pseudoscalar Higgs funnel (A4) and the
    stau-coannihilation region at large $\tanb$ (A5). We also present
    a breakdown of the dark matter fine-tuning with respect to each
    parameter, and also list the electroweak
    fine-tuning.\label{t:CMSSM}}
\end{table}

We now distinguish four regions in which the CMSSM can account for the
observed relic density of dark matter. The least interesting of these
regions is the bulk region, which once again appears at low $m_0$ and
low $m_{1/2}$, but requires a Higgs lighter than current search limits
allow.  Moving on from the inaccessible bulk region, we again find the
coannihilation strip along the side of the region ruled out because
the stau is the LSP. In comparison to Fig.\ref{f:CMSSM,t10}, it is
thin and broken. This is a sampling artefact, arising because we have
extended the range of the $m_0$ scan, and the resolution is not fine
enough to resolve the full strip. In contrast to
Fig.\ref{f:CMSSM,t10}, here the strip is plotted in purple,
designating a dark matter fine-tuning in the range $\DeltaO=30-100$.

Consider now the representative point A5 and the corresponding
fine-tuning. At large $\tanb$, the mass of the $\stau_R$ runs down
more than for low $\tanb$, and therefore a larger soft mass is
necessary to avoid a $\stau$ LSP. For low $\tanb$ the $\stau$
coannihilation region occurred when $m_0\ll m_{1/2}$, whereas here
$m_0$ and $m_{1/2}$ are much closer in magnitude. This reduces the
dominance of $m_{1/2}$ in the running and restores the need for a
precise balance of $m_0$ and $m_{1/2}$ to keep the $\stau-\neut$ mass
difference small. However, it is the influence of $\tanb$ over the
running that dominates the sensitivity.

The second region of interest shows up as two diagonal bands starting
beside the light green region ruled out because the stau is the
LSP. These lines lie to either side of the line along which
$2m_{\neut}=m_A$, and neutralino annihilation proceeds via the
resonant production of a pseudoscalar Higgs boson. This region is
known as the pseudoscalar Higgs funnel region. The WMAP lines lie
along either side of this resonance, where there is just enough of an
enhancement in the annihilation cross section to allow a bino LSP to
account for the observed relic density. As a result, the dark matter
fine-tuning price of such a region is large. This is reflected in the
purple and blue shading of the funnel region, corresponding to a dark
matter fine-tuning in the range $\DeltaO=30-300$.

We consider the representative point A4 and break down the
fine-tuning. The mass of the pseudoscalar Higgs is dependent upon the
details of the running of the soft Higgs mass-squared terms, which also
determine the Higgs vevs and thus $\tanb$. As we require REWSB and set
$\tanb$ at the start, we can run this chain of logic back the other
way. A fixed value of $\tanb$ requires a specific value of the soft
Higgs mass-squared terms at the low-energy scale. Thus the value of
$\tanb$ has a significant impact on the pseudoscalar Higgs boson
mass. This explains the dominant sensitivity of the pseudoscalar Higgs
funnel region to $\tanb$. This is to be contrasted with the NUHM,
which will allow access to the pseudoscalar Higgs funnel region
without large $\tanb$, so that naively one might expect a more natural
funnel region.

The final WMAP strip lies at large $m_0$ in all panels. The corner of
the parameter space at large $m_0$ and low $m_{1/2}$ (shown in light
red) is ruled out as $\mu^2<0$, signifying a failure of REWSB. Along
the perimeter of this region $\mu\approx 0$, and $\mu$ decreases
steadily as one approaches the boundary. As $\mu$ decreases, the
higgsino component of the LSP increases. Higgsino dark matter
annihilates very efficiently resulting in $\DM\ll \DMW$, whereas bino
dark matter generally gives $\DM\gg\DMW$. Along the WMAP strip, the
higgsino component of the $\neut$ is large enough that the
annihilation cross section is enhanced sufficiently for the relic
density to fit the WMAP data. This region is known as the focus-point
region.  It is sensitive to the composition of the lightest
neutralino, and so depends upon the difference between $\mu$ and $M_1$
at the electroweak scale. The region is plotted in red and purple,
with a couple of green regions at low $m_{1/2}$. This corresponds to a
dark matter fine-tuning of $\DeltaO \approx 10$ at low $m_{1/2}$,
rising to a dark matter fine-tuning of $\DeltaO\approx 60$ at large
$m_0$ and $m_{1/2}$. The kink at $m_{1/2}\approx450~GeV$, where the
tuning drops (signified by a change from purple to red) corresponds to
the top quark mass threshold where processes of the form $\neut\neut
\rightarrow t\overline{t}$ become kinematically allowed.

We take points A3 and A6 as representative points in this region and
break the dark matter fine-tuning down into its individual
elements. At point A3 the LSP is predominantly bino with a small, but
significant, higgsino component. The higgsino component is determined
by the relative size of $\mu(EW)$ to $M_1(EW)$. As we set $\mu(EW)$ by
requiring REWSB, $\mu(EW)$ is determined by the running of the soft
Higgs masses. Thus $\mu(EW)$ is sensitive to the soft Higgs mass-squared
terms at the GUT scale (set to $m_0^2$ in the CMSSM). It is also
sensitive to strongly-interacting sparticle masses through the
RGEs. This once again brings in a sensitivity to $m_0$, but also to
$M_3$ through its strong influence on the stop quark mass. Therefore
$\mu(EW)$ is sensitive to both $m_0$ and $m_{1/2}$. To achieve the
correct balance of bino and higgsino components one needs to balance
$\mu(EW)$ and $M_1(EW)$. These both depend strongly upon $m_{1/2}$,
with $\mu$ also dependent upon $m_0$. This common dependence on
$m_{1/2}$ reduces the dark matter fine-tuning below what one would expect.

At point A6 we consider the kink at the bottom of the higgsino/bino
line, where the higgsino and bino components are almost equal. This
would normally result in extremely efficient annihilation of
neutralinos in the early universe and give $\DM\ll\DMW$. However at
this point the neutralino mass is $m_{\neut}=79.6$~GeV, disallowing
the annihilation of neutralinos to Higgs, $Z$ bosons and
$t\overline{t}$. It is also on the $W$ boson threshold, suppressing
annihilation to $W$s. With these annihilation channels ruled out, the
normally efficient annihilation of an LSP with a substantial higgsino
fraction is suppressed sufficiently to fit the WMAP relic density. As
with point A3, the determining factors for the annihilation cross
section are the relative sizes of $\mu(EW)$ and $M_1(EW)$, resulting
in a similar pattern and magnitude of dark matter fine-tuning measures.

Once again we also calculate the electroweak fine-tuning across the
plane and plot the corresponding contours of $\DeltaEW$. As in the
case of Fig.~\ref{f:CMSSM,t10}, the degree of electroweak tuning
increases steadily with increasing $m_{1/2}$. This is because the
dominant soft term is $M_3$, the gluino mass. As $m_{1/2}$ increases,
the gluino mass and squark masses increase, requiring more precise
cancellations to reproduce the $W$ mass, and thus greater
electroweak fine-tuning.

\begin{table}[ht!]
  \begin{center}
    \begin{tabular}{|l|c|}
      \hline
      {\bf Region} & {\bf Tuning Range} \\
      \hline
      Pseudoscalar Higgs Funnel     & 60-1200+ \\
      $\stau-\neut$ coannihilation (large $m_0,~m_{1/2}$ or $\tanb$)  & 30-60\\
      $\stau-\neut$ coannihilation (low $m_0,~m_{1/2}$ and $\tanb$)  & 3-10\\
      bino/higgsino region          & 10-60\\
      \hline
    \end{tabular}
  \end{center}
  \vskip -0.5cm \caption{\small Here we summarize the DM annihilation
    channels present within the CMSSM and their associated
    tunings. This provides a reference to which we will compare the
    regions accessible within the NUHM.\label{t:CMSSMTunings}}
\end{table}

Now that we have considered the different dark matter regions present
within the CMSSM, we summarize the typical tunings in each case. We
distinguish in Table \ref{t:CMSSMTunings} the principal dark matter
regions present within the CMSSM and the corresponding amounts of
dark matter fine-tuning.
With this as our starting point, we now consider the case
of the NUHM model in which the universality between the soft Higgs
masses and the soft sfermion masses is broken.

\section{The NUHM}
\label{NUHM}

We now consider the MSSM with non-universal Higgs soft masses
(NUHM) \cite{oldnuhm,Ellis:2002wv,hep-ph/0210205}.
After breaking the universality between the soft Higgs and
sfermion masses of the CMSSM we have the following independent inputs:
\begin{equation}
  a_{NUHM}=\left\{m_0,~m_{H_1},~m_{H_2},~m_{1/2},~A_0,~\tanb,
  ~\text{sign}(\mu)\right\},
\label{NUHMPar}
\end{equation}
where $m_{H_1}$ and $m_{H_2}$ characterize the independent soft Higgs
masses.  These are subject to constraints arising from vacuum
stability and cosmological considerations, and may be negative.  As
long as $m_{H_{1,2}}^2 + \mu^2 > 0$ at the GUT scale, there is no
dangerous high-scale vacuum state, but specifying the precise
boundaries of the NUHM parameter space lies beyond the scope of this
work.

As with the CMSSM, the NUHM contains a finite number of distinct
regions in which it can provide the observed dark matter relic
density, which were catalogued in~\cite{hep-ph/0210205}. Here we
follow the approach of this previous work and reproduce the same
regions of the parameter space. The plots we present here show the
updated parameter space for the current world average for the top
mass, $m_t=170.9$~GeV, and include the current dark matter and $\gmu$
constraints. However, the primary goal of this work is rather to
analyse the fine-tuning of the dark-matter regions of the NUHM. To
this end we calculate and plot the dark-matter fine-tuning in the
allowed parameter space, and also make some comments on the amount of
electroweak fine-tuning.

As the NUHM contains the CMSSM as a limiting case, all the dark-matter
regions present in the CMSSM are present in the NUHM. In addition,
there are four new regions that are not present in the CMSSM:

\begin{itemize}
\item A pseudoscalar Higgs funnel at low $\tanb$.
\item A bulk region where $\neut$ annihilation is dominantly mediated
  via $t$-channel $\stau$ exchange which does not violate Higgs mass
  bounds.
\item A $\sneut-\neut$ coannihilation region.
\item A mixed bino/higgsino region at low $m_0$.
\end{itemize}

We shall be particularly interested in understanding how finely tuned
the NUHM parameters must be in each of these new regions.

\subsection{Comparison with the CMSSM}

The NUHM contains all the points in the CMSSM parameter
space. Therefore, we start by studying the tuning of the dark matter
points A1-6, presented in Tables~\ref{t:CMSSM,t10},\ref{t:CMSSM}, with
respect to the parameters of the NUHM.

\begin{table}[ht!]
\begin{center}
  \begin{tabular}{|l|l|l|l|l|l|l|l|l|}
      \hline
      \multicolumn{1}{|c|}{Parameter} &
      \multicolumn{2}{|c|}{A1} &
      \multicolumn{2}{|c|}{A2} &
      \multicolumn{2}{|c|}{A3} &
      \multicolumn{2}{|c|}{A4}\\
      \cline{2-9}
                  & value & $\DeltaO$ & value & $\DeltaO$ & value & $\DeltaO$ &
value & $\DeltaO$ \\
      \hline
      $m_0$       & 60    & 0.62 & 100    & 5.7 & 2030   & 200  & 540    & 8.1  \\
      $m_{H_1}$   & 60    & 0.017& 100    & 0.26& 2030   & 14   & 540    & 28   \\
      $m_{H_2}$   & 60    & 0.014& 100    & 0.26& 2030   & 230  & 540    & 30   \\
      $m_{1/2}$   & 200   & 0.99 & 500    & 5.8 & 500    & 18   & 600    & 8.0  \\
      $\tanb$     & 10    & 0.13 & 10     & 1.5 & 50     & 2.0  & 50     & 76   \\
      \hline
      $\Delta_\Omega$  &  & 0.99 &        & 5.8 &        & 230  &        & 76   \\
      \hline
      $\Delta_{EW}$    &  & 37   &        & 190 &        & 1300 &        & 230  \\
      \hline
    \end{tabular}
    \begin{tabular}{|l|l|l|l|l|}
      \hline
      \multicolumn{1}{|c|}{Parameter} &
      \multicolumn{2}{|c|}{A5} &
      \multicolumn{2}{|c|}{A6}\\
      \cline{2-5}
                  & value & $\DeltaO$ & value & $\DeltaO$  \\
      \hline
      $m_0$       & 277    & 23  & 1400   & 230  \\
      $m_{H_1}$   & 277    & 1.5 & 1400   & 10  \\
      $m_{H_2}$   & 277    & 2.5 & 1400   & 73  \\
      $m_{1/2}$   & 350    & 12  & 250    & 22  \\
      $\tanb$     & 50     & 48  & 50     & 8.2 \\
      \hline
      $\Delta_\Omega$ &    & 48  &        & 230 \\
      \hline
      $\Delta_{EW}$   &    & 92  &        & 600 \\
      \hline
    \end{tabular}
    \end{center}\vskip -0.5cm
  \caption{\small A re-analysis of the representative points A1-6 from
    Figs.~\ref{f:CMSSM,t10},\ref{f:CMSSM}, calculating their tunings
    with respect the NUHM rather than the CMSSM.\label{t:CMSSMComp}}
\end{table}

We show the dark matter
fine-tuning of these points with respect to the parameters
$a_{NUHM}$ in Table~\ref{t:CMSSMComp}. Point A1 represents the bulk
region of the CMSSM, which is inaccessible because the Higgs is
light. The primary annihilation channel is $t$-channel slepton
exchange, and the sensitivity in the CMSSM is primarily due to $m_0$
and $m_{1/2}$ as they determine the neutralino and slepton
masses. This is also true in the NUHM, and the sensitivities to the
Higgs soft masses are negligible.

Point A2 represents the low-$\tanb$ coannihilation region of the
CMSSM, in which the primary sensitivities were to $m_{1/2}$ and $m_0$,
as these determine the stau mass and the neutralino mass. Once again,
this picture changes very little in the NUHM, with the sensitivity to
the soft Higgs masses being negligible.

Points A3-6 have large $\tanb$. We recall that A3 and A6 lie in the
higgsino-bino focus-point region. In the CMSSM the primary
sensitivities were to $m_0$ and $m_{1/2}$, as $m_0$ (and to a lesser
extent $M_3$) determine the size of $\mu$, and $m_{1/2}$ determines
$M_1(EW)$. Therefore these two parameters determine the mass and
composition of the lightest neutralino, and the total CMSSM dark matter
fine-tuning of the point in the CMSSM was $\DeltaO=27$. In the NUHM we
have a very different picture. Here the total dark matter fine-tuning is
$\DeltaO=230$, and the primary sensitivities are to $m_0$ and
$m_{H_2}$. This can be explained by the process of radiative
electroweak symmetry breaking. For electroweak symmetry breaking to
occur, the Higgs (mass)$^2$ must become negative. By requiring this
process to give the correct electroweak boson masses we set the size
of $\mu$, and thus the magnitude of the higgsino component of the
lightest neutralino. Therefore to understand the sensitivity of a
higgsino-bino dark matter region, we must look for the terms that
contribute to the running of the Higgs mass-squared. First there is the
soft Higgs mass at the GUT scale, and then there are the running
effects, primarily the contribution from the stop mass. In the CMSSM,
these two terms are coupled, reducing the dependence on either one
individually. Therefore even though the scalar masses are large, the
sensitivity of $\mu$ to $m_0$ remains small. In the NUHM there is no
connection between the soft sfermion masses and the soft Higgs masses,
therefore the sensitivity returns. Therefore one should not expect
natural bino-higgsino dark matter at large $m_0$ in the NUHM. The
significant increase in the electroweak fine-tuning for these points
is due to exactly the same physics.

Points A4 and A5 represent the pseudoscalar Higgs funnel and the
stau-coannihilation band. At this value of $\tanb$, the primary
sensitivity is to $\tanb$, a feature not altered by breaking the
universality amongst the scalars.

\subsection{Detour: RGE behaviour with negative masses-squared}

To understand the dependence of the dark matter phenomenology on the
NUHM GUT scale parameters we need to understand how the soft Higgs
masses affect the RGEs, and through them the low-energy
parameters. Four low-energy parameters in particular are useful to
consider when we talk about dark matter: $\mu$, $m_A$, and
$\stau_{L,R}$. The higgsino component of the LSP is determined by
$\mu$, $m_A$ determines the position of the pseudoscalar Higgs funnel,
and the lightest stau (a mixture of $\stau_{L,R}$) mediates the
prevalent t-channel slepton exchange annihilation diagrams and
determines the efficiency of $\stau$ coannihilation channels.

After EW symmetry breaking we can write $\mu$ as:
\begin{equation}
  \mu^2=\frac{m_{H_1}^2-m_{H_2}^2\tan^2\beta}{\tan^2\beta-1} -
  \frac{1}{2}m_Z^2.
\label{e:musq}
\end{equation}
Clearly $\mu$ depends on the soft Higgs mass-squared terms and $\tanb$, as
well as other soft parameters through the RGEs. It is also useful to
consider the limit of large $\tanb$ where we can approximate
(\ref{e:musq}) as:
\begin{equation}
\mu^2=-m_{H_2}^2+\frac{m_{H_1}^2}{\tan^2\beta},
\end{equation}
assuming $|m_{H_{1,2}}^2|\gg m_Z^2$. Therefore for large $\tanb$, to
achieve REWSB and have $\mu^2>0$ we require either negative
$m_{H_2}^2$, or very large positive $m_{H_1}^2$.

The pseudoscalar Higgs mass is determined after EWSB by the relation:
\begin{equation}
  m_A^2=m_{H_1}^2+m_{H_2}^2+2\mu^2.
\label{mAsq}
\end{equation}
Clearly $m_A^2$ is strongly dependent upon the soft Higgs mass-squared
terms, $\tanb$ through its effect on $\mu$, and other soft terms
through their influence on the Higgs RGEs.

We now consider the explicit form of the soft Higgs mass-squared RGEs:
\begin{eqnarray}
\nonumber \frac{d(m_{H_1}^2)}{dt}&=&\frac{1}{8\pi^2}
\left(-3g_2^2M_2^2-g_1^2M_1^2+h_\tau^2 (m_{\stau_L}^2 +
m_{\stau_R}^2 + m_{H_1}^2 + A_\tau^2) \right.\\
&&\left.+3h_b^2(m_{\tilde{b}_L}^2 +
m_{\tilde{b}_R}^2+m_{H_1}^2+A_b^2) - 2S\right),\\
\frac{d(m_{H_2}^2)}{dt}&=&\frac{1}{8\pi^2} \left(-3g_2^2M_2^2-g_1^2M_1^2
+3h_t^2(m_{\tilde{t}_L}^2 + m_{\tilde{t}_R}^2 + m_{H_2}^2 +
A_t^2) + 2S\right),
\end{eqnarray}
where $S$ is definedly:
\begin{eqnarray}
\nonumber S&\equiv&\frac{g_1^2}{4}\left(m_{H_2}^2-m_{H_1}^2 +
2\left(m_{\tilde{Q}_L}^2-m_{\tilde{L}_L}^2-2m_{\tilde{u}_R}^2 +
m_{\tilde{d}_R}^2+m_{\tilde{e}_R}^2\right)\right.\\
&&+\left.\left(m_{\tilde{Q}_{3L}}^2-m_{\tilde{L}_{3L}}^2 -
2m_{\tilde{t}_R}^2 + m_{\tilde{b}_R}^2+m_{\stau_R}^2\right)\right).
\end{eqnarray}

The only parameters in these RGEs that we are not free to set at the
GUT scale are the Yukawa couplings $h_i$. These are set by the
requirement that the Higgs vevs should give the correct SM particle
masses:
\begin{equation}
m_{\tau,b}=\frac{1}{\sqrt{2}}h_{\tau,b} v_1,
~m_{t}=\frac{1}{\sqrt{2}}h_{t} v_2.
\end{equation}
Therefore $\tanb$ influences the RGEs indirectly through its
determination of the size of the Yukawa couplings. The Yukawa
couplings multiply the contribution to the RGEs from the soft squark
and slepton mass-squared terms and the soft Higgs terms. Therefore varying
the Yukawa couplings has a large impact on the running. As we increase
$\tanb$, we increase $v_2$ with respect to $v_1$, and so we must
decrease $h_t$ and increase $h_{\tau,b}$. Therefore we reduce the
Yukawa contribution to the running of $m_{H_2}^2$, while increasing
the contribution to the running of $m_{H_1}^2$.

Now consider the RGEs for the right and left handed stau masses:
\begin{eqnarray}
\frac{d(m_{\tilde{L}_{3L}}^2)}{dt}&=&\frac{1}{8\pi^2}
\left(-3g_2^2M_2^2-g_1^2M_1^2+h_\tau^2 \left(m_{\tilde{L}_{3L}}^2 +
m_{\stau_R}^2+m_{H_1}^2+A_\tau^2\right) -2S\right)\\
\frac{d(m_{\stau_R}^2)}{dt}&=&\frac{1}{8\pi^2} \left(-4g_1^2M_1^2 +
2h_\tau^2 \left(m_{\tilde{L}_{3L}}^2 +
m_{\stau_R}^2+m_{H_1}^2+A_\tau^2\right)+4S\right)
\end{eqnarray}
In both cases $m_{H_1}^2$ provides a substantial contribution to the
running, with a coefficient of $h_\tau$. As we have seen, increasing
$\tanb$ increases $h_\tau$ and thus increases the impact of the Higgs
masses on the running of the staus. Therefore we expect any effects of
non-universal soft Higgs masses on the stau running to be amplified
for large $\tanb$. In the CMSSM, $m_{H_1}^2$ will remain positive from
the GUT scale to the EW scale. Indeed, it is harder to achieve REWSB
if $m_{H_1}^2$ runs negative. Therefore generally this term provides a
positive contribution to both the left and right handed stau RGE and
acts to suppress the stau masses.

In the CMSSM this poses a problem. As we increase $\tanb$ we must
increase the soft stau mass to avoid it becoming the LSP. However as
we increase $m_0$ we are also increasing $m_{H_1}^2$, and thus
increase the effect on the running. This can be avoided in the
NUHM. We can set $m_{H_1}^2$ small and so avoid a very light
$\stau$.

However, there is another more subtle effect. The interaction of the
neutralinos with the stau also depends upon the composition of the
lightest stau which is determined by the mixing between
$\stau_{L,R}$. This mixing is increased if the two states are close in
mass. In the CMSSM $S$ is negligible and so $d(m_{\stau_R}^2)/dt \gg
d(m_{\tilde{L}_{3L}}^2)/dt$, resulting in the right handed stau always
being considerably lighter than the left-handed stau. In the NUHM we
can avoid this by having a large negative $S$. This acts to suppress
the left handed stau mass while increasing the right handed stau
mass. As we increase the component of the left-handed stau, we
increase the annihilation rate of neutralinos via t-channel stau
exchange.

\subsection{Sample $(m_0,m_{1/2})$ planes in the NUHM}

\begin{figure}[ht!]
  \begin{center}
    \scalebox{1.0}{\includegraphics{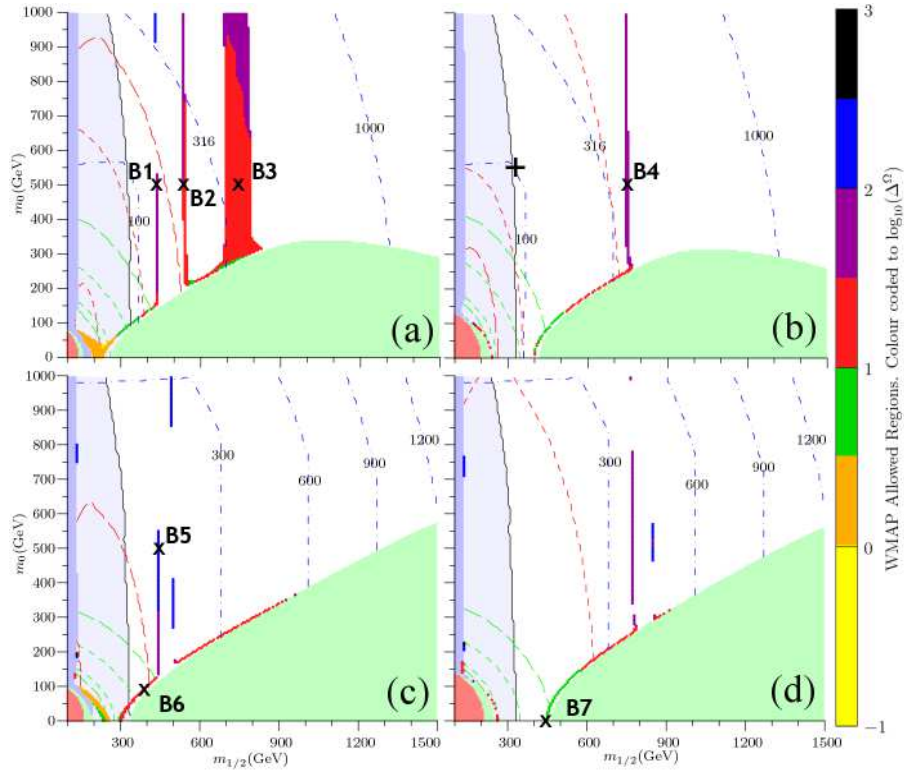}}
  \end{center}
  \vskip -0.5cm \caption{\small The $(m_0,~m_{1/2})$ plane of the NUHM
  parameter space with $A_0=0$, $\tanb=10$ and sign$(\mu)$ +ve. The
  values of $\mu$ and $m_A$ vary between the panels: (a)
  $\mu=400$~GeV, $m_A=400$~GeV, (b) $\mu=400$~GeV, $m_A=700$~GeV, (c)
  $\mu=700$~GeV, $m_A=400$~GeV, (d) $\mu=700$~GeV, $m_A=700$~GeV. This
  figure can be compared directly to Fig.~2
  in~\protect\cite{hep-ph/0210205}. The Roman cross in panel (b)
  indicates the single point where the parameter space makes contact
  with the CMSSM.\label{f:m0,m12,t,10}}
\end{figure}

Having analysed the CMSSM points from the perspective of the NUHM, we
now turn to a sampling of the full NUHM parameter space. In
Fig.~\ref{f:m0,m12,t,10}, we show $(m_{1/2}, m_0)$ planes for
$\tanb=10$, $A_0=0$ and sign$(\mu)$ positive. We set the
electroweak scale parameters $\mu$ and $m_A$ to different discrete
values in each panel as explained in the figure caption.

As we saw in the previous section, $\mu$ and $m_A$ are not high-scale
inputs into the theory, rather they are the low-energy numbers derived
from a given set of the true input parameters. However, displaying
results as functions of these parameters can be more informative. As
both have a strong dependence on $m_{H_{1,2}}^2$, we can fit a
particular value of $\mu$, $m_A$ with the correct choice of
$m_{H_{1,2}}^2$ at the GUT scale. Therefore we use a code that varies
$m_{H_{1,2}}^2$ across the parameter space to fit the designated
low-energy values of $\mu$ and $m_A$. All fine-tunings are calculated
in terms of the inputs of the NUHM as listed in (\ref{NUHMPar}).

By starting with $(m_0, m_{1/2})$ planes, we make contact with the
parameter space of the CMSSM as displayed in
Figs.~\ref{f:CMSSM,t10},~\ref{f:CMSSM}~\footnote{ We note in panel (b)
of Fig.~\ref{f:m0,m12,t,10} the appearance of a CMSSM point, the only
point in any of these planes where full GUT-scale universality is
recovered.}.  As before, low $m_0$ is ruled out by a $\stau$ LSP
(light green), and low $m_{1/2}$ results in a Higgs with $m_h<
111$~GeV (light grey with black boundary). As before, $\gmu$ favours
light sleptons, and thus low $m_0$ and $m_{1/2}$.

\begin{table}[ht!]
\begin{center}
  \begin{tabular}{|l|l|l|l|l|l|l|l|l|}
      \hline
      \multicolumn{1}{|c|}{Parameter} &
      \multicolumn{2}{|c|}{B1} &
      \multicolumn{2}{|c|}{B2} &
      \multicolumn{2}{|c|}{B3} &
      \multicolumn{2}{|c|}{B4}\\
      \cline{2-9}
                  & value & $\DeltaO$ & value & $\DeltaO$ & value & $\DeltaO$ &
value & $\DeltaO$ \\
      \hline
      $m_0$       & 500   & 32   & 500    & 8.6 & 500    & 4.6  & 500    & 12   \\
      $m_{H_1}^2$ & -80249& 16   & -126930& 12  & -248480& 0.61 & 90625  & 2.3  \\
      $m_{H_2}^2$ & 461380& 62   & 675760 & 25  & 1202900& 24   & 1194100& 60   \\
      $m_{1/2}$   & 435   & 39   & 540    & 19  & 750    & 18   & 750    & 38   \\
      $\tanb$     & 10    & 5    & 10     & 3.2 & 10     & 1.1  & 10     & 2.9  \\
      \hline
      $\Delta_\Omega$  &  & 62   &        & 25  &        & 24   &        & 60   \\
      \hline
      $\Delta_{EW}$    &  & 150  &        & 220 &        & 390  &        & 390  \\
      \hline
      \hline
      $\mu$       & 400   & -    & 400 & -   & 400 & -   & 400 & -   \\
      $m_A$       & 400   & -    & 400 & -   & 400 & -   & 700 & -   \\
      \hline
    \end{tabular}
    \\
    \begin{tabular}{|l|l|l|l|l|l|l|}
      \hline
      \multicolumn{1}{|c|}{Parameter} &
      \multicolumn{2}{|c|}{B5} &
      \multicolumn{2}{|c|}{B6} &
      \multicolumn{2}{|c|}{B7}\\
      \cline{2-7}
                  & value & $\DeltaO$ & value & $\DeltaO$ & value & $\DeltaO$ \\
      \hline
      $m_0$       & 500    & 40  & 100    & 6.3 & 0      & 0   \\
      $m_{H_1}^2$ & -416350& 110 & -400510& 12  & -79656 & 2.2 \\
      $m_{H_2}^2$ & -24320 & 4.1 & -332200& 10  & -266010& 7.4 \\
      $m_{1/2}$   & 442    & 52  & 400    & 3.5 & 445    & 4.3 \\
      $\tanb$     & 10     & 5.8 & 10     & 0.55& 10     & 1.9 \\
      \hline
      $\Delta_\Omega$ &    & 110 &        & 12  &        & 7.4 \\
      \hline
      $\Delta_{EW}$   &    & 250 &        & 250 &        & 250 \\
      \hline
      \hline
      $\mu$       & 700 & -   & 700 & -   & 700 & -   \\
      $m_A$       & 400 & -   & 400 & -   & 700 & -   \\
      \hline
    \end{tabular}
    \end{center}\vskip -0.5cm
  \caption{\small Analysis of the points B1-7, shown in
    Fig.~\ref{f:m0,m12,t,10}, which are representative of the
    pseudoscalar Higgs funnel (B1,2,4,5), mixed bino-higgsino dark
    matter (B3) and $\stau$ coannihilation regions (B6,7). We present a
    breakdown of the dark matter fine-tuning with respect to each parameter
    of the NUHM. We give the value of $m_{H_{1,2}}^2$, but the tunings
    are calculated with respect to $m_{H_{1,2}}$.\label{t:m0,m12,t,10}}
\end{table}

The dark matter phenomenology shows some similarities to and some
marked differences from the CMSSM. First, we see a familiar $\stau$
coannihilation region alongside the region with a $\stau$ LSP. As in
the CMSSM, this region is plotted in red and green, designating a
tuning of $\DeltaO=3-30$. The only new feature of the coannihilation
region here is that effects of the non-universal Higgs soft masses
alter the running of the stau mass, which allows access to regions
with $m_0=0$. We can access $m_0=0$ with small negative $m_{H_1}^2$
and larger negative $m_{H_2}^2$. The combination of a small Yukawa
contribution (due to low $\tanb$ along with small $|m_{H_1}^2|$) along
with negative $S$ results in the stau mass that increases as we run
down from the GUT scale, allowing an acceptable stau mass even with
$m_0=0$.

The points B6 and B7 are representative of the $\stau$ coannihilation
region, and the breakdowns of their tunings are also shown in
Table~\ref{t:m0,m12,t,10}. The dependences on $m_0$ and $m_{1/2}$ are
similar to what was observed in the CMSSM. However, the dominant
sensitivities are now to $m_{H_{1,2}}$. For both these points the soft
Higgs mass-squared terms are large and negative at the GUT scale. As we
have seen, these soft parameters have a significant effect on the stau
RGE. Therefore the coannihilation strip exhibits tuning with respect to
these parameters. The total sensitivity remains low suggesting that,
even though the soft Higgs masses play a role in the running, the
dominant contribution to the stau mass is still from $M_1$.

More distinctive deviations from the familiar CMSSM phenomenology
arise in the forms of the strong vertical dark matter regions at
particular values of $m_{1/2}$. In panel (a) three vertical strips are
present. To understand these lines we need to consider the mass and
composition of the lightest neutralino. The bino component of the
lightest neutralino is determined by $M_1(EW)\approx 0.4
m_{1/2}(GUT)$, whereas the wino component is determined by
$M_2(EW)\approx 0.8 m_{1/2}(GUT)$. Hence, $M_2(EW)>M_1(EW)$ throughout
the NUHM parameter space, and we never have a large wino component in
the LSP. Of more importance is the higgsino component, determined by
$\mu(EW)$. When $\mu(EW)\approx M_1(EW)$, there will be a sizeable
higgsino component in the LSP. In panel (a) we have set $\mu=400$~GeV
and $m_A=400$~GeV. Therefore, when $m_{1/2}\approx 1000$~GeV,
$M_1(EW)\approx \mu$ and the lightest neutralino will be a
bino/higgsino mixture.  However, for $m_{1/2}\gg 1000$~GeV the
lightest neutralino is mainly a higgsino, with a mass
$m_{\neut}\approx 400$~GeV, whereas for $m_{1/2}\ll 1000$~GeV the
$\neut$ is predominantly a bino and has a mass determined by
$M_1(EW)$.

With this in mind, we can understand the vertical lines in panel (a)
at particular values of $m_{1/2}$. At $m_{1/2}=500$~GeV, the lightest
neutralino is a bino with a mass $m_{\neut}\approx 200$~GeV. As the
pseudoscalar Higgs mass is $m_A=400$~GeV throughout, this results in
resonant neutralino annihilation through the pseudoscalar Higgs. As a
result, the relic density is below the WMAP value across the region
$450<m_{1/2}<530$. On the edges of the resonance the relic density may
fall within the narrow range favoured by astrophysics.  The lines are
mostly plotted in purple and blue, corresponding to a large degree of
dark matter fine-tuning $\DeltaO=30-300$.
However, it is interesting that the edge
of the resonance at larger $m_{1/2}$ is plotted in red. This is the
first instance of an acceptable pseudoscalar Higgs resonance region
with relatively low dark-matter fine-tuning, thanks to the larger
higgsino fraction in the LSP at larger $m_{1/2}$. Both the LSP mass
and the mass of $m_A$ are sensitive to $\mu$, so the mass of the LSP
and $m_A$ are coupled. This mitigates the dark matter
tuning of the pseudoscalar
Higgs funnel to an extent.

Points B1 and B2 lie on either side of the pseudoscalar Higgs funnel,
where it is interesting to break the dark matter
fine-tuning measure down into its
component parts. Unsurprisingly, the dark matter
fine-tuning is due to a balancing
act between the mass of the pseudoscalar Higgs, as shown by the large
sensitivity to $m_{H_2}$, and the neutralino mass, as shown by the
sensitivity to $m_{1/2}$.

The other vertical strip is a wide region around $m_{1/2}\approx
750$~GeV, where the predominantly bino LSP acquires a sufficient
higgsino admixture to suppress the relic density to the observed
value. This wide band is plotted in red and purple, representing a
tuning of $10-30$, similar to that of the low-$m_0$ end of the
bino/higgsino region within the CMSSM. There is also a thin line of
green at the base of this band where it meets the coannihilation
strip. This shows that the interaction of a stau coannihilation region
with a bino/higgsino LSP reduces the overall tuning of either region
alone.

Point B3 is representative of this region. The primary dark matter
fine-tunings are clearly with respect to $m_{H_2}$ and $m_{1/2}$. In
this case, the dark matter fine-tuning is related to the balance
between the roles of these terms in determining $\mu$ and $M_1$ at the
electroweak scale.  This is the first instance of a mixed
bino-higgsino region at low $m_0$ that we find in the NUHM. It is
therefore interesting to see that, even with a TeV-scale value of
$m_{H_2}$, the tuning remains relatively small. This is in contrast to
the point A3, at which a TeV-scale soft Higgs mass gave large
fine-tuning. To understand the origin of both electroweak and dark
matter fine-tunings, it is useful to consider the analytic form of the
dependence of the low-energy parameter $\mu(EW)$ on the GUT-scale soft
inputs. For $\tanb=10$, we have:
\begin{eqnarray}
  \nn \frac{m_Z^2}{2}=&&-0.94\mu^2+0.010m_{H_1}^2-0.19M_2^2
  -0.0017M_1^2-0.63m_{H_2}^2+0.38m_{Q_3}^2\\
  \nn &&+0.38m_{U_3}^2
  +0.093A_t^2-0.011A_tM_1-0.070A_tM_2-0.30A_tM_3\\
  &&+2.51M_3^2+0.0059M_1M_2+0.028M_1M_3+0.195M_2M_3 ,
  \label{musq}
\end{eqnarray}
from which we can see that, when $m_{Q_3}^2=m_{H_{1,2}}^2$, the terms
from the scalars approximately cancel. This explains the jump in
sensitivity when this universality is broken. On the other hand, in
order to obtain a small value of $\mu$, these soft scalar terms need
to provide a large contribution to balance out the contribution from
$M_3^2$. This explains why we find a higgsino/bino only at large $m_0$
in the CMSSM. In the NUHM, it is unnecessary to go to large $m_0$,
just large $m_{H_2}^2$. By keeping $m_0$, and thus $m_{Q_3}^2$, small
one keeps the dark matter and electroweak fine-tunings associated with
these parameters under control. The remaining large electroweak
fine-tuning associated with point B3 is due to $M_{1/2}$ being quite
sizeable.

In panel (b) we have $\mu=400$~GeV and $m_A=700$~GeV. Here we see only
the lower edge of the pseudoscalar resonance. This band lies at
$750$~GeV and is plotted in purple, once again showing the large dark
matter fine-tuning we expect of such resonances. The upper edge of the
resonance never appears, because at larger $m_{1/2}$ the lightest
neutralino is dominantly higgsino. Therefore, at values of $m_{1/2}$
above the resonance, the higgsino nature of the LSP suppresses the
relic density so that the relic density never rises enough to fit the
WMAP measurement.

Point B4 illustrates the component dark matter fine-tunings at the
resonance. As with point B1, the dark matter fine-tuning is large, and
primarily due to sensitivities to $m_{H_2}^2$ and $m_{1/2}$. These are
due to their effects on the pseudoscalar Higgs mass and the mass of
the LSP, respectively.

In panel (c) we set $\mu=700$~GeV and $m_A=400$~GeV. As the
pseudoscalar mass is the same, the pseudoscalar Higgs funnel is
centred in the same place as in panel (a). However, the higgsino
fraction of the LSP has dropped, increasing the overall relic
density. Therefore one must go closer to the resonance before the
enhancement to the annihilation cross section is sufficient to fit the
observed relic density. Therefore the WMAP lines are closer to the
peak of the resonance, and require more precise dark matter
fine-tuning than for the lower value of $\mu$. This is shown by
examining the component dark matter fine-tunings for point B5. The
dark matter fine-tuning is large, and primarily due to $m_{H_2}$
through its influence on the pseudoscalar Higgs mass. In panel (c)
there is no region in which the LSP is higgsino, due to the larger
value of $\mu$. One would have to go to $m_{1/2}\approx 1700$~GeV
before the LSP acquires a significant higgsino fraction as
$\mu=700$~GeV.

Finally, in panel (d) we take $\mu=700$~GeV and $m_A=700$~GeV, and we
see the pseudoscalar Higgs resonance at $m_{1/2}=800$~GeV as expected,
and it remains finely tuned, as before. As in panel (c), the value of
$\mu$ is too large to find a region of the parameter space in which
the neutralino is higgsino.

Overall, Fig.~\ref{f:m0,m12,t,10} shows some similarities and some
marked deviations from the phenomenology of the CMSSM. A $\stau$
coannihilation band appears in roughly the same region of the
parameter space and exhibits slightly larger tuning. This increase is
due to the effect of the soft Higgs masses squared on the $\stau$
running. A further deviation from the CMSSM comes in the form of a
pseudoscalar Higgs funnel at low $\tanb$. In most cases this exhibits
dark matter fine-tuning similar to the CMSSM, supporting the
observation that resonances require significant dark matter
fine-tuning wherever they appear. The exception is when the LSP has a
significant higgsino fraction. The sensitivity of both $m_A$ and
$m_{\neut}$ to $\mu$ lowers the required dark matter fine-tuning
significantly.  There also appears a band of higgsino/bino dark matter
and we find that the tuning is of the same order as in the CMSSM,
which is surprising. In the CMSSM the focus-point region has
relatively low dark matter fine-tuning because of the cancellation
among the scalar masses. By breaking this universality, this
cancellation is broken. However, it also allows us to access mixed
bino/higgsino regions at low $m_0$. This enhances other annihilation
channels, such as that via $t$-channel slepton exchange, and we
require a smaller higgsino admixture to obtain a suitable dark matter
relic density. These regions recover the relatively low dark matter
fine-tuning of the focus-point region of the CMSSM.

Finally, we note that we have also calculated the electroweak
fine-tuning across these planes in the NUHM parameter space, and find
it to be very similar to that of the CMSSM planes studied previously.

\begin{figure}[ht!]
  \begin{center}
    \scalebox{1.0}{\includegraphics{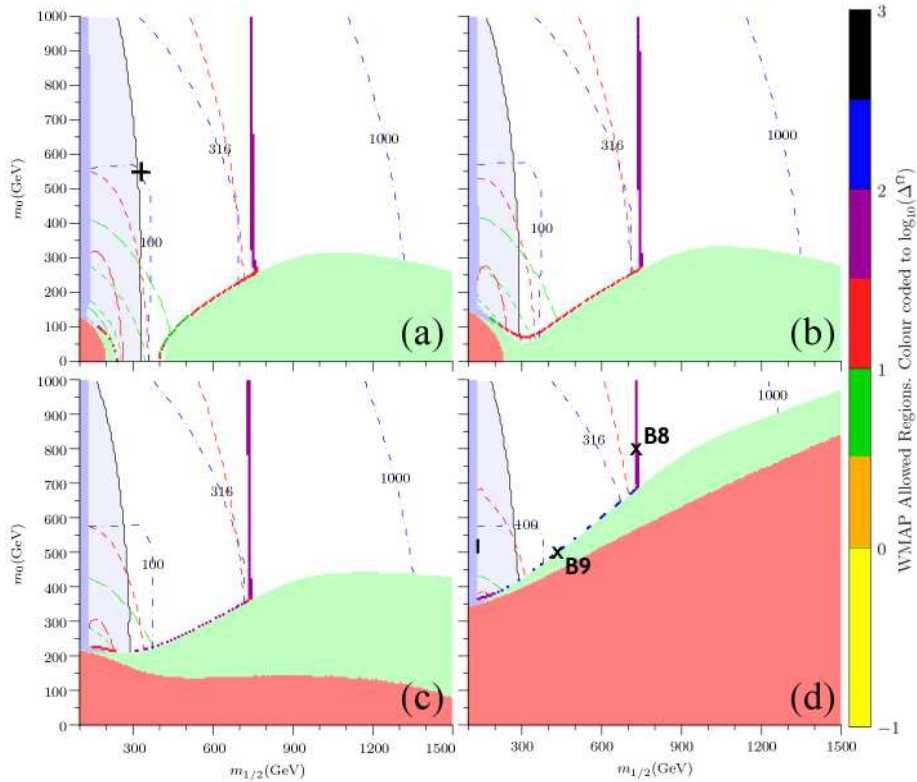}}
  \end{center}
  \vskip -0.5cm \caption{\small Sample $(m_{1/2}, m_0)$ planes of the
  NUHM parameter space with $A_0=0$, $\mu=400$~GeV, $m_A=700$~GeV and
  sign$(\mu)$ and the following values of $\tanb$: (a) $\tanb=10$, (b)
  $\tanb=20$, (c) $\tanb=35$, (d) $\tanb=50$. The Roman cross in panel
  (a) shows the single point where the NUHM makes contact with the
  MSSM. \label{f:mu,400,mA,700}}
\end{figure}

We now consider the behaviours of these regions as $\tanb$
increases. In Fig.~\ref{f:mu,400,mA,700} we take $\mu=400$~GeV,
$m_A=700$~GeV, $A_0=0$ and sign($\mu$) positive, and in panels (a)-(d)
we take successively larger values of $\tanb$~\footnote{ Note the
CMSSM point in panel (a) of Fig.~\ref{f:mu,400,mA,700}.}.

\begin{table}[ht!]
  \begin{center}
  \begin{tabular}{|l|l|l|l|l|}
      \hline
      \multicolumn{1}{|c|}{Parameter} &
      \multicolumn{2}{|c|}{B8} &
      \multicolumn{2}{|c|}{B9}\\
      \cline{2-5}
                  & value & $\DeltaO$ & value & $\DeltaO$ \\
      \hline
      $m_0$       & 800     & 21   & 500    & 150 \\
      $m_{H_1}^2$ & 1609600 & 30   & 892230 & 98  \\
      $m_{H_2}^2$ & 1357100 & 75   & 379620 & 1.3 \\
      $m_{1/2}$   & 730     & 14   & 432    & 35  \\
      $\tanb$     & 50      & 50   & 50     & 290 \\
      \hline
      $\Delta_\Omega$  &  & 75   &        & 290 \\
      \hline
      $\Delta_{EW}$    &  & 420  &        & 130 \\
      \hline
      \hline
      $\mu$       & 400   & -    & 400 & -   \\
      $m_A$       & 700   & -    & 700 & -   \\
      \hline
    \end{tabular}
  \end{center}\vskip -0.5cm
  \caption{\small Points B8 and B9, shown in
    Fig.~\ref{f:mu,400,mA,700} exemplify the behaviour of the
    pseudoscalar Higgs funnel (B8) and the stau-coannihilation region
    (B9) at large $\tanb$ within the NUHM. We present a breakdown of
    the dark matter fine-tuning with respect to each parameter of the
    NUHM. We give the value of $m_{H_{1,2}}^2$, although the tunings
    are calculated with respect to
    $m_{H_{1,2}}$.\label{t:mu,400,mA,700}}
\end{table}

As the value of $\tanb$ increases, the mass and composition of the
neutralino across the $(m_{1/2}, m_0)$ plane remains essentially
unaltered. Therefore we find the lower edge of the pseudoscalar funnel
in the same place in all panels. Studying the breakdown of the dark matter
fine-tunings of the pseudoscalar Higgs funnel for large $\tanb$ at
point B8, we find that in this case all parameters show a significant
tuning. The primary tunings come from $m_{H_2}$ and $\tanb$. From
(\ref{mAsq}),(\ref{e:musq}) we can see that this comes from the
dominant sensitivity of $m_A^2$, through $\mu$, on both $\tanb$ and
$m_{H_2}^2$.

The pseudoscalar Higgs funnel remains in the same place as $\tanb$
increases, whereas the $\stau$-coannihilation strip moves
considerably. For larger $\tanb$ the Yukawa contribution to the stau
running is enhanced. This suppresses the stau mass, requiring larger
$m_0$ to avoid a stau LSP. For very low $m_0$ there is a brick red
region. This is not due to a failure of REWSB, but rather it is due to
a tachyonic stau mass.

By increasing $\tanb$ we reduce the dominance of $M_1$ in the running
of the stau mass-squared. This breaks the link between the masses of the
$\stau$ and the $\neut$, resulting in an increase of the tuning
required for the coannihilation region at larger values of
$\tanb$. Point B9 is representative of the $\stau$ coannihilation
region at large $\tanb$. Though there is large sensitivity to the soft
gaugino mass $m_{1/2}$, this is eclipsed by the sensitivities to
$\tanb$ and $m_0$.

\subsection{Sample $(\mu,m_A)$ planes}

\begin{figure}[ht!]
  \begin{center}
    \scalebox{1.0}{\includegraphics{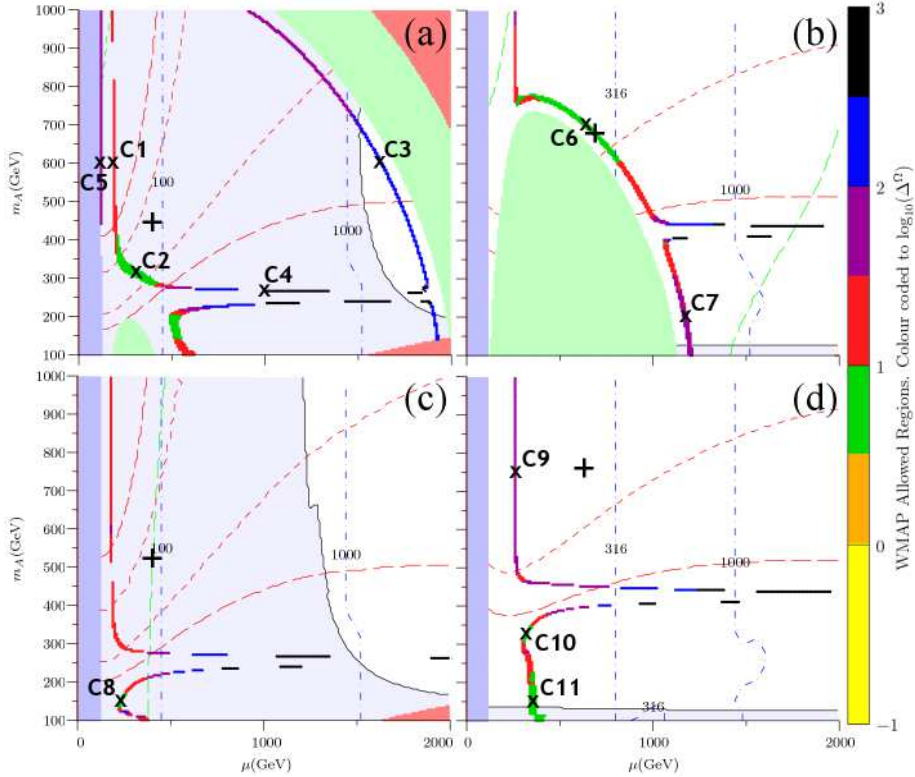}}
  \end{center}
  \vskip -0.5cm \caption{\small Sample NUHM $(\mu,~m_A)$ planes with
$A_0=0$, $\tanb=10$ and sign$(\mu)$ positive, and different values of
$m_0$ and $m_{1/2}$: (a) $m_0=100$~GeV, $m_{1/2}=300$~GeV, (b)
$m_0=100$~GeV, $m_{1/2}=500$~GeV, (c) $m_0=300$~GeV,
$m_{1/2}=300$~GeV, (d) $m_0=300$~GeV, $m_{1/2}=500$~GeV. The Roman
crosses in each panel show where the NUHM meets the
CMSSM.\label{f:mu,mA,t,10}}
\end{figure}

We now discuss other planes so as to explore more features of the NUHM
model. In Fig.~\ref{f:mu,mA,t,10} we consider $(\mu,~m_A)$ planes with
$A_0=0$, $\tanb=10$ and sign$(\mu)$ positive, taking different values of
$m_0$ and $m_{1/2}$ in the various panels. As before, though we plot
results in terms of $\mu$ and $m_A$, the primary variables are
$m_{H_{1,2}}^2$.

We first consider some overall features. Low $\mu$ is ruled out in all
cases by the appearance of a light chargino (light blue). In panels (a)
and (c), the low value of $m_{1/2}=300~\text{GeV}$ results in a Higgs
boson with $m_h<111$~GeV across much of the parameter space. This bound is
very sensitive to variations in the top mass, so regions that are excluded
here could be allowed with a higher top mass. On the other hand, in panels
(b) and (d), only very low values of $m_A$ result in a problematically
light Higgs. In panels (a) and (b), the low value of $m_0=100~\text{GeV}$
results in a light $\stau$. This gives a region at low $\mu$ and $m_A$ in
which the $\stau$ is the LSP and as such is ruled out (light green).
Finally, in panel (a) $m_{1/2}$ and $m_0$ are light enough to give light
sneutrinos. For large $\mu$ and $m_A$, one of the sneutrinos becomes the
LSP, ruling out this corner of the parameter space~\footnote{Sneutrinos
are massive, neutral and weakly-interacting, and so could in principle
account for the dark matter. However, they generally give $\DM\ll\DMW$.
This can be avoided in some models with right-handed neutrinos, but this
possibility lies beyond the scope of the MSSM, so we do not consider it
further here.}.

\begin{table}[ht!]
  \begin{center}
  \begin{tabular}{|l|l|l|l|l|l|l|l|l|}
      \hline
      \multicolumn{1}{|c|}{Parameter} &
      \multicolumn{2}{|c|}{C1}&
      \multicolumn{2}{|c|}{C2}&
      \multicolumn{2}{|c|}{C3}&
      \multicolumn{2}{|c|}{C4}\\
      \cline{2-9}
                  & value & $\DeltaO$ & value & $\DeltaO$ & value & $\DeltaO$ &
value & $\DeltaO$ \\
      \hline
      $m_0$       & 100     & 0.99 & 100    & 4.7  & 100     & 13   & 100      & 2.7 \\
      $m_{H_1}^2$ & 279530  & 0.38 & -27130 & 0.99 & -2289300& 79   & -966510  & 390 \\
      $m_{H_2}^2$ & 188070  & 17   & 111220 & 3.5  & -3637400& 110  & -1248800 & 330 \\
      $m_{1/2}$   & 300     & 16   & 300    & 0.20 & 300     & 30   & 300      & 46  \\
      $\tanb$     & 10      & 0.8  & 10     & 0.51 & 10      & 0.57 & 10       & 1.4 \\
      \hline
      $\Delta_\Omega$  &    & 17   &        & 4.7  &         & 110  &          & 390 \\
      \hline
      $\Delta_{EW}$    &    & 74   &        & 73   &         & 1300 &          & 500 \\
      \hline
      \hline
      $\mu$       & 190   & -    & 300 & -   & 1620  & -    & 1000& -   \\
      $m_A$       & 600   & -    & 315 & -   & 600   & -    & 268 & -   \\
      \hline
    \end{tabular}
  \begin{tabular}{|l|l|l|l|l|l|l|l|l|}
      \hline
      \multicolumn{1}{|c|}{Parameter} &
      \multicolumn{2}{|c|}{C5}&
      \multicolumn{2}{|c|}{C6}&
      \multicolumn{2}{|c|}{C7}&
      \multicolumn{2}{|c|}{C8}\\
      \cline{2-9}
                  & value & $\DeltaO$ & value & $\DeltaO$ & value & $\DeltaO$ &
value & $\DeltaO$ \\
      \hline
      $m_0$       & 100     & 3.3  & 100    & 5.6  & 100     & 4.7  & 300      & 0.64\\
      $m_{H_1}^2$ & 319380  & 2.7  & -22696 & 0.59 & -1445500& 35   & -70487   & 6.7 \\
      $m_{H_2}^2$ & 222330  & 76   & -25802 & 0.68 & -1440300& 34   & 246160   & 2.4 \\
      $m_{1/2}$   & 300     & 85   & 500    & 5.6  & 500     & 3.4  & 300      & 3.6 \\
      $\tanb$     & 10      & 2.7  & 10     & 1.3  & 10      & 0.068& 10       & 1.2 \\
      \hline
      $\Delta_\Omega$  &    & 85   &        & 5.6  &         & 35   &          & 6.7 \\
      \hline
      $\Delta_{EW}$    &    & 74   &        & 210  &         & 590  &          & 79  \\
      \hline
      \hline
      $\mu$       & 120   & -    & 640 & -   & 1170  & -    & 235 & -   \\
      $m_A$       & 600   & -    & 700 & -   & 200   & -    & 150 & -   \\
      \hline
    \end{tabular}
  \begin{tabular}{|l|l|l|l|l|l|l|l|l|}
      \hline
      \multicolumn{1}{|c|}{Parameter} &
      \multicolumn{2}{|c|}{C9}&
      \multicolumn{2}{|c|}{C10}&
      \multicolumn{2}{|c|}{C11}\\
      \cline{2-7}
                  & value & $\DeltaO$ & value & $\DeltaO$ & value & $\DeltaO$ \\
      \hline
      $m_0$       & 300     & 6.1  & 300    & 0.55 & 300     & 1.3  \\
      $m_{H_1}^2$ & 401250  & 0.62 & -106330& 5.1  & -211970 & 0.067\\
      $m_{H_2}^2$ & 568170  & 40   & 525900 & 2.5  & 497010  & 6.8  \\
      $m_{1/2}$   & 500     & 31   & 500    & 0.023& 500     & 6.5  \\
      $\tanb$     & 10      & 1.5  & 10     & 1.0  & 10      & 0.30 \\
      \hline
      $\Delta_\Omega$  &    & 40   &        & 5.1  &         & 6.8  \\
      \hline
      $\Delta_{EW}$    &    & 180  &        & 180  &         & 180  \\
      \hline
      \hline
      $\mu$       & 260   & -    & 320 & -   & 350   & -    \\
      $m_A$       & 750   & -    & 325 & -   & 150   & -    \\
      \hline
    \end{tabular}
  \end{center}\vskip -0.5cm
  \caption{\small Points C1-11, shown in Fig.~\ref{f:mu,mA,t,10},
    illustrate the behaviour of mixed bino-higgsino dark matter
    (C1,5,9), sneutrino coannihilation (C3), the pseudoscalar Higgs
    funnel (C4), stau-coannihilation (C2,6,7,11), and the
    pseudoscalar funnel region with a mixed bino-higgsino LSP
    (C8,10). We present breakdowns of the dark matter fine-tuning with
    respect to each parameter of the NUHM. We give the value of
    $m_{H_{1,2}}^2$, but the fine-tunings are calculated with respect
    to $m_{H_{1,2}}$.\label{t:mu,mA,t,10}}
\end{table}

Before considering the dark matter regions, we first note the
composition of the lightest neutralino in different regions of the
parameter space. In panels (a) and (c) $M_1(EW)\approx 120$~GeV,
whereas in panels (b) and (d) $M_1(EW)\approx 200$~GeV. Therefore for
$\mu\gg 120(200)$~GeV the LSP is predominantly a bino with a mass
$m_{\neut}\approx 120(200)$~GeV. Below these values of $\mu$, the
lightest neutralino acquires a significant higgsino fraction.

As before, the regions that fit the relic density favoured by WMAP are
displayed with the corresponding dark matter fine-tuning colours.  The
vertical dark matter band at low $\mu$ in all panels features a mixed
bino/higgsino dark matter particle, and is plotted in red and purple,
corresponding to a dark matter fine-tuning $\DeltaO\approx 20-40$.
Points C1 and C9 are representative of the higgsino/bino band, and
their dark matter fine-tunings are broken down in
Table~\ref{t:mu,mA,t,10}. As one would expect, the fine-tuning is due
to a balancing act between $m_{H_2}$ and $m_{1/2}$, as these determine
the composition of the lightest neutralino, and thus its annihilation
rate.

Point C5 also represents a bino/higgsino region. It is unusual in
that, normally, the larger the higgsino component, the more efficient
the annihilation. However, in this case, as $\mu$ is lowered from
point C1, the higgsino fraction increases and yet the dark matter
relic density rises again. Indeed, C5 is almost evenly split between
higgsino and bino components and yet it fits the WMAP relic density
measurement. This is due to the fact that, as $\mu$ drops, the mass of
the neutralino is lowered. At point C5 the LSP becomes lighter than
the $Z$, shutting off the annihilation channel $\neut \neut
\rightarrow ZZ$. This reduces the annihilation cross section enough
that the relic density is acceptable once more.

Another region that is easy to pick out is the pseudoscalar Higgs
funnel along $m_A=240(400)$~GeV for $m_{1/2}=300(500)$~GeV
respectively. As before, these funnels require significant dark matter
fine-tuning and as such are predominantly plotted in blue and black,
showing a tuning $\DeltaO>100$. Point C4 is a representative point
whose dark matter fine-tuning breakdown we display in
Table~\ref{t:mu,mA,t,10}. In previous pseudoscalar Higgs regions the
dark matter fine-tuning was due primarily to $m_{H_{1,2}}$ and
$m_{1/2}$. Here we find that the sensitivity to the Higgs masses has
increased significantly. From (\ref{mAsq}) this is easy to
understand. If we increase $\mu$ while keeping $m_A$ the same we must
carefully balance the large $m_{H_{1,2}}^2$ contributions to give the
required $m_A$. This careful balancing manifests as a steadily
increasing sensitivity of $m_A$ to the Higgs soft masses as we
increase $\mu$. This translates to a large sensitivity of the
pseudoscalar Higgs funnel.

At the other end of the spectrum, there is a region of the
pseudoscalar Higgs funnel at low $\mu$ with remarkably low
tuning. This occurs when there is a significant higgsino fraction in
the LSP, such as at points C8 and C10. In this region, both $m_A$ and
the neutralino mass are sensitive to $\mu$. This results in the mass
of the neutralino and the pseudoscalar being coupled, and reduces the
sensitivity of the mass difference $\Delta_m=m_A-2m_{\neut}$. At
points C8 and C10 the dominant annihilation channels are to heavy
quarks via an $s$-channel pseudoscalar Higgs. Remarkably the total
dark matter fine-tunings of the points are only 6.7 and 5.1
respectively.

As the $\tilde{\nu}_{e,\mu}$ become the LSPs in the large $\mu$, large
$m_A$ region of panel (a), there is a corresponding sneutrino
coannihilation region lying parallel to its boundary, which is plotted
in purple and blue indicating a dark matter fine-tuning
$\DeltaO>80$. Point C3 is a representative of this region, whose dark
matter fine-tuning breakdown is also displayed in Table
\ref{t:mu,mA,t,10}. The dark matter fine-tuning is large, and comes
primarily from the Higgs sector. It is the existence of large negative
$m_{H_1}^2$ that allows for light sneutrinos. Thus the sneutrino
masses are very sensitive to the Higgs soft mass-squared parameters, and
this is reflected in the dark matter fine-tuning. There is also some
dark matter fine-tuning with respect to $m_{1/2}$ that is typical of
the need to balance the bino mass against that of a coannihilation
partner with an uncorrelated mass.

Finally, the light $\stau$ at low $\mu$ and $m_0$ has an effect on the
dark matter relic density. As the mass of the $\stau$ is reduced, the
annihilation cross section is increased via $t$-channel slepton
exchange. Also, as one approaches the region in which the stau is the
LSP, there are additional contributions from $\stau-\neut$
coannihilation processes. These two effects combine to give dark
matter bands along the edges of the stau LSP region in panels (a) and
(b). Points C2 in panel (a) and C6,7 in panel (b) are representative
points. At point C2 the annihilation proceeds through equal parts of
$t$-channel $\sel,~\smu,~\stau$ annihilation (15-20\% each),
annihilation to $b,\overline{b}$ via off-shell pseudoscalar Higgs
bosons (18\%) and $\stau$ coannihilation (15\%). Only the
coannihilation processes would be expected to exhibit a high
sensitivity to the soft parameters, as $t$-channel processes are
fairly insensitive and the point is far from the pseudoscalar
resonance, reducing significantly the sensitivity of the $s$-channel
pseudoscalar process. As a result, we have a region that arises from a
mixture of channels and exhibits low tuning. The subdominant role of
coannihilation explains why there is so little dark matter fine-tuning
with respect to $m_{1/2}$. The role of the stau in both the
coannihilation and $t$-channel processes explains the dominant dark
matter fine-tuning with respect to $m_0$, and the dependence on
$m_{H_2}$ appears from running effects.

Unfortunately, point C2 results in a light Higgs with $m_{h}=110$~GeV,
which is probably unacceptably low, even allowing for the theoretical
uncertainty in the calculation of its mass. On the other hand, panel
(b) has a larger value of $m_{1/2}$ and hence Higgs mass.  However,
the masses of the LSP and the sleptons are also increased. This
decreases the slepton $t$-channel annihilation cross sections,
requiring larger contributions from processes that are finely tuned in
order to fit the WMAP relic density, which is apparent at points C6
and C7~\footnote{ We note that there is a CMSSM point very close
to C6}. At point C6, $t$-channel slepton annihilation only accounts
for 3\% of the annihilation rate via each channel (9\% overall). The
remaining 91\% is made up entirely of coannihilation processes,
dominantly with $\stau$, but also $\sel,\smu$. As this plane has low
$m_0$, the slepton masses are predominantly determined by
$m_{1/2}$. Once again, there is the familiar pattern of dark matter
fine-tunings for a low-$\tanb$, low-$m_0$ slepton coannihilation
region. The overall dark matter fine-tuning is low, and what
fine-tuning does exists is due to $m_0$ and $m_{1/2}$. Point C7 tells
a slightly different story. The pattern of annihilation channels is
almost identical to C6, and we see the typical dark matter
fine-tunings of a coannihilation region in the sensitivity to $m_0$
and $m_{1/2}$. However, the dark matter fine-tuning with respect to
$m_{H_{1,2}}$ has increased dramatically, due to the massive increase
in $m_{H_{1,2}}^2$ between points C6 and C7. Now the stau running is
dominated by the Higgs mass-squared terms rather than the gaugino
mass, and the coannihilation region becomes fine-tuned once again.

There is one further interesting region. In panel (d) at low $m_A$
there is a kink in the higgsino/bino region. The band moves to larger
$\mu$ and the dark matter fine-tuning drops dramatically. The band is
plotted in green rather than purple, indicating a dark matter
fine-tuning of less than 10. The kink in the band appears at
$m_A=280~$GeV. Around this region the LSP is predominantly a bino with
a small but significant higgsino component, and the LSP has a mass of
around 200~GeV. As the pseudoscalar mass drops, the masses of the
heavy Higgs, $H$, and the charged Higgses, $H^\pm$, also
decrease. Around $m_A=280$~GeV, the annihilation channels
$\neut\neut\rightarrow hA,W^\pm H^\mp,ZH$ open up, which are
kinematically forbidden at larger $m_A$. These can progress through
either $t$-channel neutralino (chargino) exchange or $s$-channel Higgs
and Z processes. They require a small higgsino component, but
significantly less than the higgsino/bino region represented by point
C9. This balance of the higgsino and bino components of the LSP
appears in the sensitivity of point C11 on $m_{1/2}$ and
$m_{H_2}$. Thus C11 represents a higgsino/bino region with low dark
matter fine-tuning - something that does not exist in the CMSSM. This
is because a large negative $m_{H_1}^2$ is needed to achieve low
$m_{A,H,H^\pm}$.

\begin{figure}[ht!]
  \begin{center}
    \scalebox{1.0}{\includegraphics{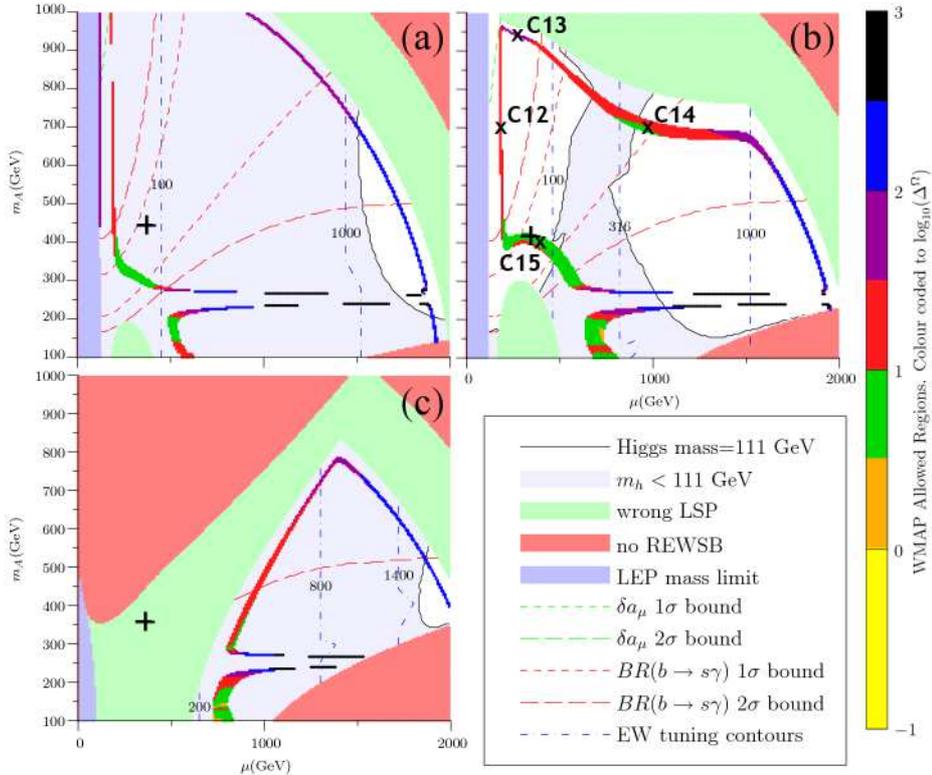}}
  \end{center}
  \vskip -0.5cm \caption{\small Sample NUHM $(\mu,~m_A)$ planes with
  $A_0=0$, $m_0=100$~GeV, $m_{1/2}=300$~GeV and sign$(\mu)$ positive,
  and the following values of $\tanb$: (a) $\tanb=10$, (b) $\tanb=20$,
  (c) $\tanb=35$. We do not show a plot for $\tanb=50$ as the
  parameter space is entirely excluded. The Roman crosses in each
  panel show where the NUHM meets the CMSSM.\label{f:m0,100,m12,300}}
\end{figure}

We now consider in Fig.~\ref{f:m0,100,m12,300} the behaviours of these
regions as $\tanb$ increases. We have set $m_0=100$~GeV,
$m_{1/2}=300$~GeV, $A_0=0$ and increase $\tanb$ in steps in each
panel.

We note first the bulk features of the plane. As noted previously,
increasing $\tanb$ decreases the mass of the lightest stau. Thus plots
at larger $\tanb$ have larger regions ruled out because the LSP is a
$\stau$, and we do not show very large $\tanb$ because at $\tanb=50$
the stau mass becomes tachyonic across the entire plane. By $\tanb=35$
the light stau rules out all the parameter space below
$\mu=200$~GeV. The mass of the light Higgs is also sensitive to
$\tanb$, and is in all cases very close to $m_h=111$~GeV, so it only
takes a small shift to cause a significant change in the area plotted
in light grey.

\begin{table}[ht!]
  \begin{center}
  \begin{tabular}{|l|l|l|l|l|l|l|l|l|}
      \hline
      \multicolumn{1}{|c|}{Parameter} &
      \multicolumn{2}{|c|}{C12}&
      \multicolumn{2}{|c|}{C13}&
      \multicolumn{2}{|c|}{C14}&
      \multicolumn{2}{|c|}{C15}\\
      \cline{2-9}
                  & value & $\DeltaO$ & value & $\DeltaO$ & value & $\DeltaO$ &
value & $\DeltaO$ \\
      \hline
      $m_0$       & 100     & 1.0  & 100    & 15   & 100     & 7.2  & 100    & 8.7 \\
      $m_{H_1}^2$ & 477840  & 1.4  & 858150 & 48   & -532000 & 6.5  & -14857 & 0.28\\
      $m_{H_2}^2$ & 175680  & 17   & 96420  & 6.3  & -1263800& 12   & -2379  & 0.041\\
      $m_{1/2}$   & 300     & 16   & 300    & 32   & 300     & 10   & 300    & 4.6  \\
      $\tanb$     & 20      & 0.56 & 20     & 21   & 20      & 8.7  & 20     & 6.5  \\
      \hline
      $\Delta_\Omega$  &    & 17   &        & 48   &         & 12   &        & 8.7  \\
      \hline
      $\Delta_{EW}$    &    & 71   &        & 71   &         & 480  &        & 78   \\
      \hline
      \hline
      $\mu$       & 185   & -    & 275 & -   & 1000  & -    & 400 & -   \\
      $m_A$       & 700   & -    & 940 & -   & 700   & -    & 400 & -   \\
      \hline
    \end{tabular}
  \end{center}\vskip -0.5cm
  \caption{\small Points C12-15, shown in Fig.~\ref{f:m0,100,m12,300},
    are representative of the higgsino/bino region (C12), the
    sneutrino coannihilation region (C13) and the
    stau-coannihilation/bulk region (C15,14) with increasing $\tanb$
    within the NUHM. We present a breakdown of the dark matter
    fine-tuning with respect to each parameter of the NUHM. We give
    the value of $m_{H_{1,2}}^2$ but the tunings are calculated with
    respect to $m_{H_{1,2}}$.\label{t:m0,100,m12,300}}
\end{table}

The most significant change in the dark matter phenomenology is due to
the varying $\stau$ mass. Between panels (a) and (b) the stau
bulk/coannihilation region moves to larger $\mu$ and $m_A$. We also
find a significant stau region at large $m_A$. These features are
represented by points C15 and C14 respectively~\footnote{ We note that
there is a CMSSM point very close to C15.}. Comparing C15 directly
to C2, we see from Table~\ref{t:m0,100,m12,300} that the dark matter
fine-tuning is due primarily to $m_0$ and $\tanb$. This is because
these parameters determine the mass of the lighter stau and this is
the primary source of sensitivity for bulk regions. There is also a
degree of sensitivity to $m_{1/2}$, as there is a significant
coannihilation contribution that requires the LSP and stau mass to be
balanced. At point C14 one has similar degrees of dark matter
fine-tuning with respect to $\tanb,~m_{1/2}$ and $m_0$. However, there
is now also large dark matter fine-tuning with respect to
$m_{H_{1,2}}$, due to the larger magnitude of the soft higgsino
mass-squared terms. The stau mass in this region becomes highly
sensitive to $m_{H_2}^2$.

The other regions are little changed from before. Point C12
exemplifies the mixed bino/higgsino region at increasing $\tanb$. It
can be compared directly to point C1, and we see that the component
dark matter fine-tunings are virtually identical.  Point C13 is
representative of the sneutrino coannihilation region and can be
compared to point C3. Once again the dark matter fine-tuning is due
primarily to the soft Higgs masses through their impacts on the
running of the sneutrino masses.

\begin{figure}[ht!]
  \begin{center}
    \scalebox{1.0}{\includegraphics{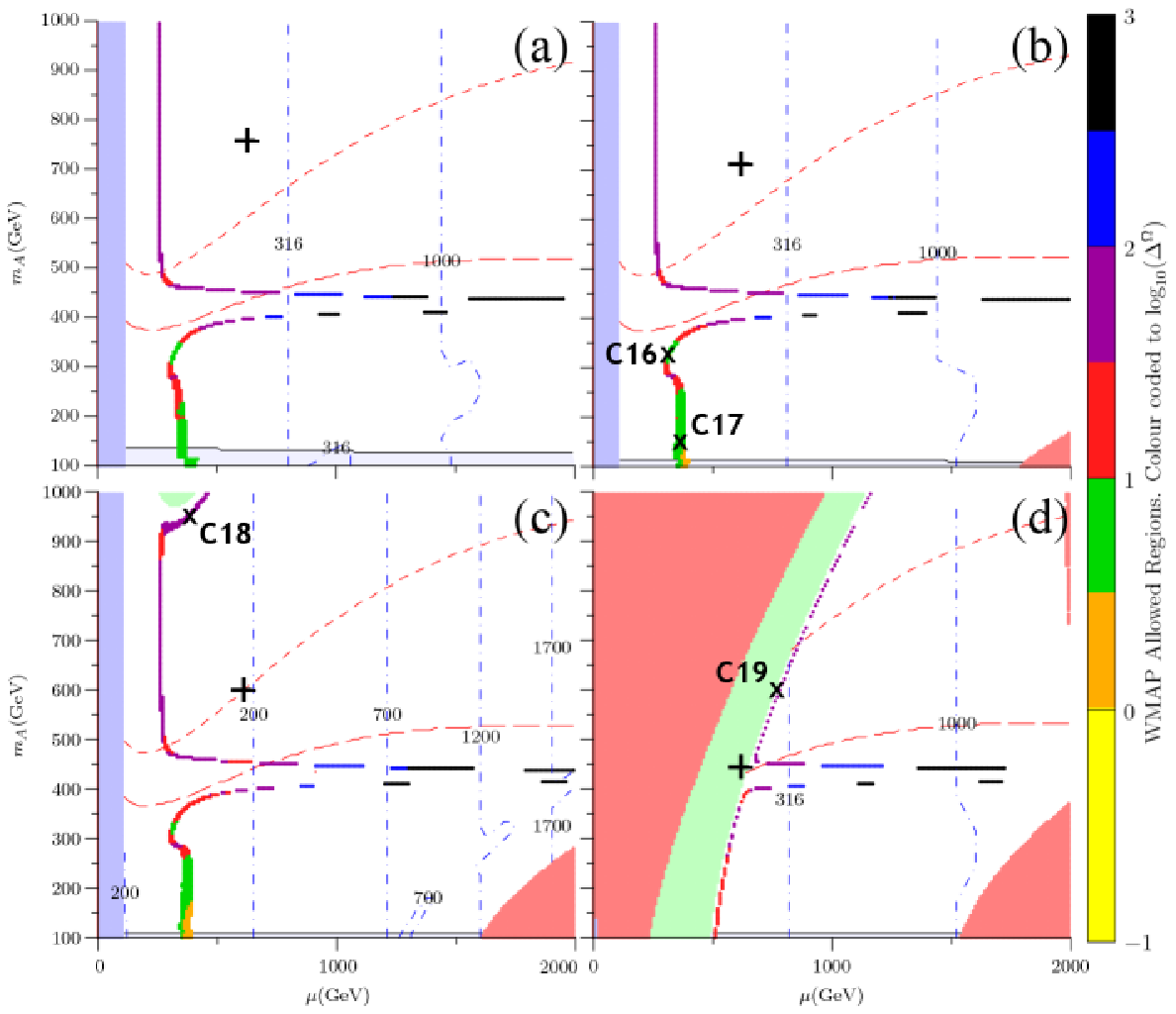}}
  \end{center}
  \vskip -0.5cm \caption{\small Sample NUHM $(\mu,~m_A)$ planes with
  $A_0=0$, $m_0=300$~GeV, $m_{1/2}=500$~GeV, sign$(\mu)$ positive and
  different values of $\tanb$: (a) $\tanb=10$, (b) $\tanb=20$, (c)
  $\tanb=35$, (d) $\tanb=50$. The Roman crosses in each panel show
  where the NUHM meets the CMSSM.\label{f:m0,300,m12,500}}
\end{figure}

Much of the low-$m_0$ parameter space is forbidden by a light Higgs
and/or a light stau. We now consider the effect of increasing $\tanb$
in a more open part of the parameter space. We take
Fig.~\ref{f:mu,mA,t,10}(d) with $m_0=300$~GeV, $m_{1/2}=500$~GeV as a
starting point and increase $\tanb$ steadily, as seen in
Fig.~\ref{f:m0,300,m12,500}.  In contrast to
Fig.~\ref{f:m0,100,m12,300}, the bulk features remain fairly stable
for moderate values of $\tanb$. The first hint of a change appears in
panel (c) at $\tanb=35$, where we see a small region at large $m_A$ in
which the stau is the LSP. This expands to cut off low $\mu$ for
$\tanb=50$.

\begin{table}[ht!]
  \begin{center}
  \begin{tabular}{|l|l|l|l|l|l|l|l|l|}
      \hline
      \multicolumn{1}{|c|}{Parameter} &
      \multicolumn{2}{|c|}{C16}&
      \multicolumn{2}{|c|}{C17}&
      \multicolumn{2}{|c|}{C18}&
      \multicolumn{2}{|c|}{C19}\\
      \cline{2-9}
                  & value & $\DeltaO$ & value & $\DeltaO$ & value & $\DeltaO$ &
value & $\DeltaO$ \\
      \hline
      $m_0$       & 300     & 1.0  & 300    & 0.94 & 300     & 52   & 300    & 37  \\
      $m_{H_1}^2$ & -47935  & 2.0  & -170090& 0.86 & 1021100 & 22   & -52957 & 4.4 \\
      $m_{H_2}^2$ & 518240  & 3.8  & 475340 & 4.2  & 390800  & 0.58 & -281880& 2.3 \\
      $m_{1/2}$   & 500     & 3.1  & 500    & 3.7  & 500     & 25   & 500    & 20  \\
      $\tanb$     & 20      & 3.0  & 20     & 0.21 & 35      & 86   & 50     & 49  \\
      \hline
      $\Delta_\Omega$  &    & 3.8  &        & 4.2  &         & 86   &        & 49  \\
      \hline
      $\Delta_{EW}$    &    & 170  &        & 170  &         & 170  &        & 290 \\
      \hline
      \hline
      $\mu$       & 315   & -    & 360 & -   & 380   & -    & 780 & -   \\
      $m_A$       & 325   & -    & 150 & -   & 950   & -    & 600 & -   \\
      \hline
    \end{tabular}
  \end{center}\vskip -0.5cm
  \caption{\small Points C16-19, shown in
    Fig.~\ref{f:m0,300,m12,500}, illustrate the behaviours of
    the mixed bino-higgsino, the pseudoscalar Higgs funnel (C16) and
    the stau-coannihilation/bulk regions (C17,18,19) at increasing
    values of $\tanb$ within the NUHM. We present a breakdown of the
    dark matter fine-tuning with respect to each parameter of the NUHM. We
    give the value of $m_{H_{1,2}}^2$, but the tunings are calculated
    with respect to $m_{H_{1,2}}$.\label{t:m0,300,m12,500}}
\end{table}

There are few dark matter surprises at larger $\tanb$. The
pseudoscalar Higgs funnel and mixed higgsino/bino regions remain
relatively unaltered throughout. The interaction of the pseudoscalar
Higgs funnel with the higgsino/bino LSP continues to provide a
favourable degree of tuning in panels (b) and (c). We take point C16
as a representative point, and break the tuning down in
Table~\ref{t:m0,300,m12,500}. As for point C10, the tuning is small
and the annihilation is primarily due to annihilation to heavy quarks
via an $s$-channel pseudoscalar Higgs.

Point C17 exemplifies the behaviour of a predominantly bino LSP with a
small higgsino admixture that can annihilate to $hA,~ZH$ and $W^\pm
H\mp$. As with point C11, the dark matter fine-tuning is small and
mostly due to the composition of the LSP, through $m_{1/2}$ and
$m_{H_2}$.

Point C18 is in the new stau coannihilation region that appears at
large $\tanb$. For $m_0=300$~GeV, $m_{1/2}=500$~GeV the staus are too
heavy to contribute significantly to $t$-channel slepton exchange, so
this region is pure coannihilation. The stau mass is mainly determined
by $m_0$ and $\tanb$, and must be balanced against a predominantly
bino LSP. Therefore, the tuning is dominated by $\tanb$ and $m_0$ with
a secondary dependence on $m_{1/2}$.  The coannihilation grows
significantly by $\tanb=50$ and point C19 represents this trend.  As
with point C18, we find the tuning to be due to $m_0$ and $\tanb$,
with a secondary dependence on $m_{1/2}$.

Throughout all of these parameter scans we have also calculated the
electroweak fine-tuning and found it to be of the same order as that
found in the CMSSM for typical scales of soft masses considered.

\subsection{Sample $(\mu,m_{1/2})$ planes}

Finally,we consider sample $(\mu,~m_{1/2})$ planes in the NUHM. These
are interesting, e.g., because $\mu$ and $m_{1/2}$ are the parameters
that determine the mass and composition of the lightest
neutralino~\footnote{Note that in the following plots $m_{1/2}$ is the
GUT-scale soft mass, whereas $\mu$ is the electroweak-scale Higgs
term. This is in contrast to the plots of~\cite{hep-ph/0210205} where
the plots were in terms of $M_2(EW)$ and $\mu(EW)$.}.

\begin{figure}[ht!]
  \begin{center}
    \scalebox{1.0}{\includegraphics{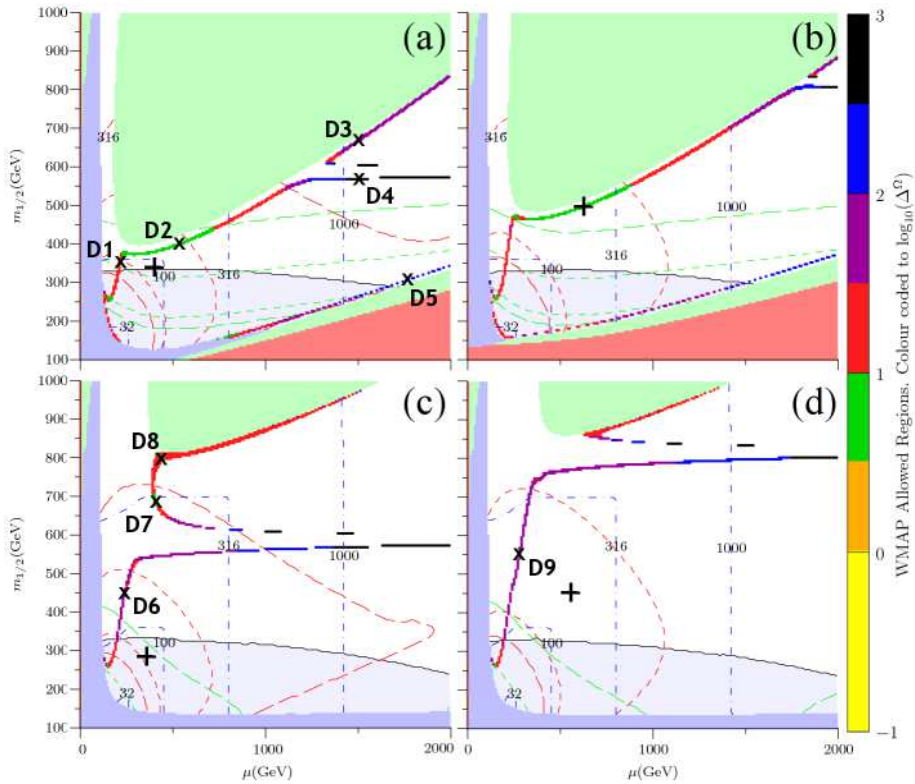}}
  \end{center}
  \vskip -0.5cm \caption{\small Sample NUHM $(\mu,~m_{1/2})$ planes
  with $A_0=0$, $\tanb=10$, sign$(\mu)$ positive and varying $m_0$
  and $m_A$: (a) $m_0=100$~GeV, $m_A=500$~GeV, (b)
  $m_0=100$~GeV, $m_A=700$~GeV, (c) $m_0=300$~GeV, $m_A=500$~GeV, (d)
  $m_0=300$~GeV, $m_A=700$~GeV. The Roman crosses in each
  panel show where the NUHM meets the CMSSM.\label{f:mu,m12,t,10}}
\end{figure}

In Fig.~\ref{f:mu,m12,t,10} we set $A_0=0$, $\tanb=10$ and take
discrete values of $m_0$ and $m_A$. We see that either low $\mu$ or
low $m_{1/2}$ results in a light chargino that violates particle
searches (light blue). Low $m_{1/2}$ also results in problems with a
light Higgs (light grey with a black boundary). On the other hand,
large $m_{1/2}$ results in a neutralino with a mass above that of the
stau (light green). The exception is low $\mu$ where the neutralino is
a higgsino and $m_{\neut}$ is insensitive to $m_{1/2}$. In panels (a)
and (b) we have $m_0=100$~GeV. This, combined with low $m_{1/2}$ and
large $\mu$ results in a region in which the LSP is a sneutrino (light
green).

\begin{table}[ht!]
  \begin{center}
  \begin{tabular}{|l|l|l|l|l|l|l|l|l|}
      \hline
      \multicolumn{1}{|c|}{Parameter} &
      \multicolumn{2}{|c|}{D1}&
      \multicolumn{2}{|c|}{D2}&
      \multicolumn{2}{|c|}{D3}&
      \multicolumn{2}{|c|}{D4}\\
      \cline{2-9}
                  & value & $\DeltaO$ & value & $\DeltaO$ & value & $\DeltaO$ &
value & $\DeltaO$ \\
      \hline
      $m_0$       & 100     & 1.0  & 100    & 7.2  & 100     & 3.6  & 100     & 1.1 \\
      $m_{H_1}^2$ & 160370  & 0.30 & -97736 & 3.3  & -2206600& 39   & -2147800& 360 \\
      $m_{H_2}^2$ & 255000  & 18   & -20502 & 0.70 & -2345100& 41   & -2588200& 300 \\
      $m_{1/2}$   & 350     & 17   & 400    & 4.7  & 670     & 5.9  & 570     & 59  \\
      $\tanb$     & 10      & 0.51 & 10     & 1.1  & 10      & 0.027& 10      & 0.070\\
      \hline
      $\Delta_\Omega$  &    & 18   &        & 7.2  &         & 41   &         & 360 \\
      \hline
      $\Delta_{EW}$    &    & 96   &        & 140  &         & 1100 &         & 1100\\
      \hline
      \hline
      $\mu$       & 210   & -    & 530 & -   & 1500  & -    & 1500& -   \\
      $m_A$       & 500   & -    & 500 & -   & 500   & -    & 500 & -   \\
      \hline
    \end{tabular}
  \begin{tabular}{|l|l|l|l|l|l|l|l|l|}
      \hline
      \multicolumn{1}{|c|}{Parameter} &
      \multicolumn{2}{|c|}{D5}&
      \multicolumn{2}{|c|}{D6}&
      \multicolumn{2}{|c|}{D7}&
      \multicolumn{2}{|c|}{D8}\\
      \cline{2-9}
                  & value & $\DeltaO$ & value & $\DeltaO$ & value & $\DeltaO$ &
value & $\DeltaO$ \\
      \hline
      $m_0$       & 100     & 13   & 300    & 7.1  & 300     & 0.38 & 300     & 7.4 \\
      $m_{H_1}^2$ & -2963700& 110  & 109920 & 0.42 & -120550 & 5.7  & -216760 & 0.91\\
      $m_{H_2}^2$ & -4438200& 140  & 490790 & 39   & 845460  & 5.1  & 1153100 & 9.0 \\
      $m_{1/2}$   & 310     & 32   & 450    & 30   & 680     & 4.7  & 800     & 2.6 \\
      $\tanb$     & 10      & 0.52 & 10     & 1.1  & 10      & 1.9  & 10      & 0.70\\
      \hline
      $\Delta_\Omega$  &    & 140  &        & 39   &         & 5.7  &         & 9.0 \\
      \hline
      $\Delta_{EW}$    &    & 1600 &        & 160  &         & 300  &         & 400 \\
      \hline
      \hline
      $\mu$       & 1785  & -    & 240 & -   & 410   & -    & 430 & -   \\
      $m_A$       & 500   & -    & 500 & -   & 500   & -    & 500 & -   \\
      \hline
    \end{tabular}
  \begin{tabular}{|l|l|l|}
      \hline
      \multicolumn{1}{|c|}{Parameter} &
      \multicolumn{2}{|c|}{D9}\\
      \cline{2-3}
                  & value & $\DeltaO$\\
      \hline
      $m_0$       & 300     & 5.6  \\
      $m_{H_1}^2$ & 291010  & 0.38 \\
      $m_{H_2}^2$ & 661620  & 42   \\
      $m_{1/2}$   & 550     & 34   \\
      $\tanb$     & 10      & 1.4  \\
      \hline
      $\Delta_\Omega$  &    & 42   \\
      \hline
      $\Delta_{EW}$    &    & 210  \\
      \hline
      \hline
      $\mu$       & 280   & -    \\
      $m_A$       & 700   & -    \\
      \hline
    \end{tabular}
  \end{center}\vskip -0.5cm
  \caption{\small Points D1-9, shown in
    Fig.~\ref{f:mu,m12,t,10}, are representative of bino-higgsino dark
    matter (D1,6,9), stau-coannihilation (D2,3,8), the pseudoscalar
    Higgs funnel (D4) and its interaction with mixed bino-higgsino
    dark matter (D7), and sneutrino coannihilation (D5). We present
    breakdowns of the dark matter fine-tuning with respect to each parameter
    of the NUHM. We give the value of $m_{H_{1,2}}^2$ but the tunings
    are calculated with respect to $m_{H_{1,2}}$.\label{t:mu,m12,t,10}}
\end{table}

We see once again the familiar dark matter regions of the previous
plots. The pseudoscalar Higgs funnel appears as a pair of horizontal
lines and exhibits large dark matter fine-tuning, and is characterized
by the point D4 in Table~\ref{t:mu,m12,t,10}. Here we see that the
large dark matter fine-tuning is due to the soft Higgs masses through
their influence on $m_A$, and to $m_{1/2}$ through its influence on
the neutralino mass.

The exception to this large dark matter fine-tuning is where the
pseudoscalar funnel interacts with a higgsino/bino LSP and there is a
small corner with low fine-tuning, as characterized by point D7. The
annihilation here is mainly to heavy quarks via an $s$-channel
pseudoscalar Higgs, and yet the total tuning is only 5.7. As noted
previously, this relatively small dark matter fine-tuning comes from
the common sensitivity of $m_A$ and $m_{\neut}$ on $\mu$.

There is also a $\stau$ coannihilation region in all four plots, which
lies alongside the region ruled out due to a stau LSP. It exhibits
similar tuning to the CMSSM. We break down the dark matter
fine-tunings of this region at points D2 and D3, finding that at both
points the tuning with respect to $m_0$ and $m_{1/2}$ is standard for
a stau coannihilation strip at low $m_0$~\footnote{This is also true
for the CMSSM point seen in panel (b).}. Point D3 has larger tuning
because this region of parameter space requires large negative soft
Higgs masses, which now dominate the determination of the mass of the
light stau.

The sneutrino coannihilation region shows up alongside the sneutrino
LSP region. Once again we find it to require significant dark matter
fine-tuning, although this decreases steadily as one moves to lower
$\mu$. Point D5 is a representative point with, as before, large dark
matter fine-tuning that depends on the soft Higgs masses.

Each plot also has a dark matter region at low $\mu$ that lies along a
diagonal in the $(\mu, m_{1/2})$ plane, incorporating points
D1,6,9. These regions are mixed bino/higgsino regions. In all cases
the pseudoscalar Higgs and heavy Higgs bosons are sufficiently massive
that annihilation of the mixed LSP proceeds mainly through the
channels $\neut \neut \rightarrow W^+W^-(ZZ)$, via $t$-channel
chargino (neutralino) exchange. This process is very sensitive to the
composition of the LSP and the masses of the exchanged
particles. Therefore there is significant dark matter fine-tuning with
respect to $m_{H_2}$ and $m_{1/2}$ at all these points.

Finally, we consider the point D8 where the coannihilation strip and
the mixed bino/higgsino strips meet. The combination of annihilation
channels has a beneficial effect, with the overall dark matter fine-tuning
dropping to 9.

\begin{figure}[ht!]
  \begin{center}
    \scalebox{1.0}{\includegraphics{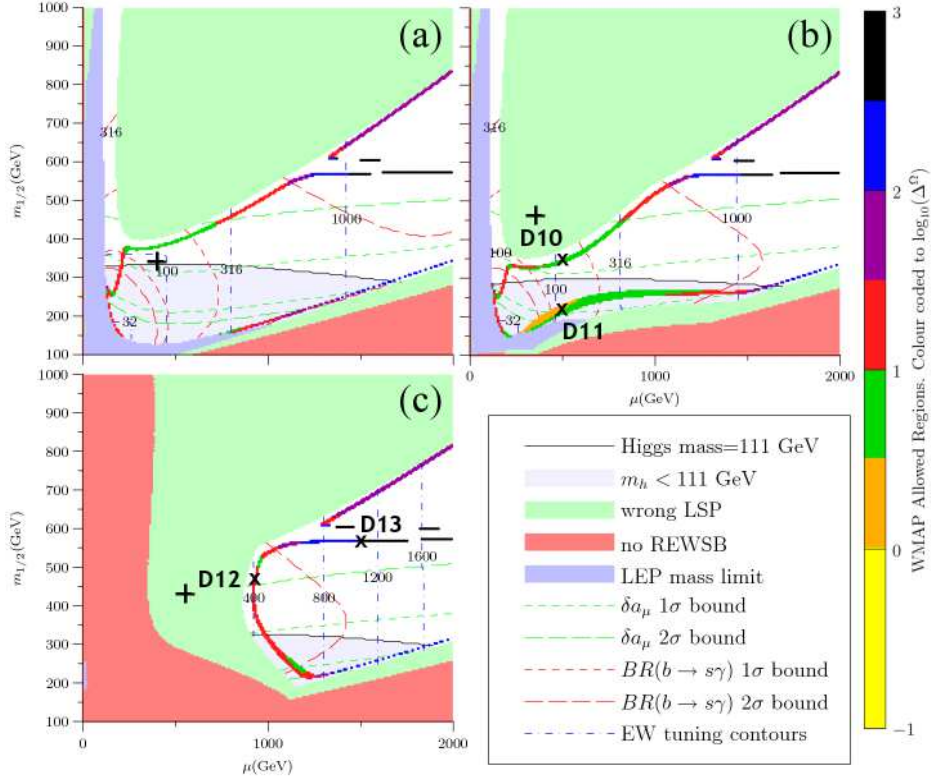}}
  \end{center}
  \vskip -0.5cm \caption{\small Sample NUHM $(\mu,~m_{1/2})$ planes
  with $A_0=0$, $m_0=100$~GeV, $m_A=500$~GeV, sign$(\mu)$ positive and
  $\tanb$ varying: (a) $\tanb=10$, (b) $\tanb=20$, (c) $\tanb=35$. We
  do not show a plane for $\tanb=50$, as this part of the parameter
  space is entirely excluded. The Roman crosses in each panel show
  where the NUHM meets the CMSSM.\label{f:m0,100,mA,500}}
\end{figure}

Once again it is interesting to go beyond $\tanb=10$, to understand
how the phenomenology changes with $\tanb$. In
Fig.~\ref{f:m0,100,mA,500} we consider $(\mu, m_{1/2})$ planes with
$m_0=100$~GeV, $m_A=500$~GeV and steadily increasing values of
$\tanb$. As we saw before, increasing $\tanb$ decreases the $\stau$
mass, causing the stau LSP regions to encroach on the parameter
space. By $\tanb=35$ the light stau rules out all values of low $\mu$.
As noted earlier, at such a low value of $m_0$, $\tanb=50$ has a
tachyonic stau and so is not shown here.

\begin{table}[ht!]
  \begin{center}
  \begin{tabular}{|l|l|l|l|l|l|l|l|l|}
      \hline
      \multicolumn{1}{|c|}{Parameter} &
      \multicolumn{2}{|c|}{D10} &
      \multicolumn{2}{|c|}{D11} &
      \multicolumn{2}{|c|}{D12} &
      \multicolumn{2}{|c|}{D13}\\
      \cline{2-9}
                  & value & $\DeltaO$ & value & $\DeltaO$ & value & $\DeltaO$ &
value & $\DeltaO$ \\
      \hline
      $m_0$       & 100     & 8.8  & 100    & 2.6  & 100     & 5.0  & 100     & 0.69\\
      $m_{H_1}^2$ & -21261  & 0.37 & -377   & 0.0  & -658190 & 15   & -2200200& 320 \\
      $m_{H_2}^2$ & -73998  & 1.5  & -243830& 0.91 & -772520 & 0.27 & -2597000& 280 \\
      $m_{1/2}$   & 345     & 4.7  & 220    & 2.1  & 470     & 4.1  & 567     & 32  \\
      $\tanb$     & 20      & 8.5  & 20     & 3.1  & 35      & 0.99 & 35      & 11  \\
      \hline
      $\Delta_\Omega$  &    & 8.8  &        & 3.1  &         & 15   &        & 320 \\
      \hline
      $\Delta_{EW}$    &    & 120  &        & 120  &         & 420  &        & 1100\\
      \hline
      \hline
      $\mu$       & 500   & -    & 500 & -   & 930   & -    & 1500& -   \\
      $m_A$       & 500   & -    & 500 & -   & 500   & -    & 500 & -   \\
      \hline
    \end{tabular}
  \end{center}\vskip -0.5cm
  \caption{\small Properties of points D10-13, shown in
    Fig.~\ref{f:m0,100,mA,500} which are representative of the
    pseudoscalar Higgs funnel (D13) and the stau-coannihilation/bulk
    region (D10,11,12) at increasing $\tanb$ within the NUHM. We
    present a breakdown of the dark matter fine-tuning with respect to
    each parameter of the NUHM. We give the value of $m_{H_{1,2}}^2$
    but the tunings are calculated with respect to
    $m_{H_{1,2}}$.\label{t:m0,100,mA,500}}
\end{table}

The change in the stau mass is the dominant factor that changes the
dark matter phenomenology. With the lighter stau, the contribution to
neutralino annihilation from $t$-channel stau exchange increases. We
consider two points D10 and D11 in panel (b). At point D10 the
annihilation is still dominated by coannihilation effects, but the
growing contribution from $t$-channel stau exchange helps to lower the
dark matter tuning. The dark matter fine-tuning is predominantly due
to $m_0$ and $\tanb$ through their influence on the mass of the
lighter stau, with a subsidiary fine-tuning with respect to
$m_{1/2}$. In contrast, point D11 lies in a dark matter band where the
annihilation of neutralinos is dominantly through $t$-channel slepton
exchange. As a result the dark matter fine-tuning is small, and due
primarily to $m_0$ and $\tanb$ through their influence on the slepton
masses.

As we move to larger $\tanb$, the coannihilation and bulk regions
meet. In panel (c) we take point D12 as a representative of the
meeting of these two regions. However, by this stage one needs large
soft Higgs mass-squared parameters and the stau mass is sensitive to
these, rather than to $m_0$ and $m_{1/2}$. Therefore there is large
fine-tuning with respect to $m_{H_1}$.  Finally, point D13 is
representative point of the pseudoscalar Higgs funnel for large
$\tanb$. As before, we find the dark matter fine-tuning to be large and
predominantly due to the soft Higgs masses. This chimes with the
general behaviour of the pseudoscalar Higgs funnel throughout our
study.

\section{Conclusions}
\label{Conc}

\begin{table}
\begin{center}
\begin{tabular}{|l|c|}
\hline
{\bf Region} & {\bf Typical $\DeltaO$} \\
\hline
$\stau$ bulk region                                          & 1-5\\
$\stau-\neut$ coannihilation                                 & 4-80\\
Bino annihilation via pseudoscalar Higgs Funnel              & 30-1200+\\
Bino/higgsino annihilation via pseudoscalar Higgs Funnel     & 3-10\\
Bino/higgsino region, $m_{\neut}>m_{H,A}$                    & 30-300\\
Bino/higgsino region, $m_{\neut}<m_{H,A}$                    & 2-10\\
$\sneut-\neut$ coannihilation                                & 15-200\\
\hline
\end{tabular}
\end{center}
\vskip -0.5cm \caption{\small A summary of the different dark matter
  regions present within the NUHM and typical values of the
  corresponding dark matter fine-tunings. We also note that
  combinations from many channels decrease the overall
  tuning. \label{t:NUHMTunings}}
\end{table}

We summarize in Table~\ref{t:NUHMTunings} the ranges of dark matter
fine-tunings found in the different dark matter regions appearing in
the NUHM, which may be compared to those found previously in the
CMSSM, as shown in Tables~\ref{t:CMSSM,t10},\ref{t:CMSSM} and
summarised in Table~\ref{t:CMSSMTunings}. Comparing first the bulk
regions, which require the smallest amounts of dark matter fine-tuning
in both the CMSSM and the NUHM, we see that CMSSM point A2 has a low
amount of dark matter fine-tuning that is at the end of the range
found in the NUHM. However, point A2 has too small a value of $m_h$,
and hence is not acceptable, whereas the NUHM can circumvent this
restriction. Thus, {\it the NUHM provides access to dark matter which
is less fine-tuned than in the CMSSM}. Turning then to the $\stau$
coannihilation regions, we see that the NUHM offers an option of lower
fine-tuning than that found in the CMSSM at point A5. For reasons
explained earlier in the text, there are very large variations in the
amounts of dark matter fine-tuning required in the pseudoscalar Higgs
funnel region of the NUHM, and the amount found at the CMSSM point A4
lies within this range. In addition, the NUHM contains a region in
which a bino/higgsino LSP annihilates via a pseudoscalar Higgs boson
that requires dramatically less fine-tuning than the pseudoscalar
funnel in the CMSSM. Likewise, NUHM analogues of the focus points A3
and A6 shown in Table~\ref{t:CMSSM} require dark matter fine-tunings
that are substantially less than in the CMSSM. One new region appears
in the NUHM that has no CMSSM analogue, namely the $\sneut-\neut$
coannihilation region.  At least in the examples studied, this
requires rather more dark matter fine-tuning than the $\stau-\neut$
coannihilation region.

Although it was not the primary focus of this paper, we have also
calculated the electroweak fine-tuning across the NUHM parameter
space, and found it to be of the same order of magnitude as in other
MSSM studies.

Generally speaking, the fact that the NUHM has more parameters offers
more possibilities to find regions with particularly small (or large)
dark matter fine-tunings. The smaller amounts of dark matter
fine-tuning generally occur in regions where several different
(co)annihilation processes contribute to the final dark matter
density, e.g., where a bino/higgsino band meets a stau coannihilation
region. Conversely, there are instances where a tight correlation is
necessary between two {\it a priori} independent MSSM parameters, such
as the stau and neutralino masses along a coannihilation strip, which is
imposed automatically in the lower-dimensional parameter space of the
CMSSM, resulting in smaller amount of dark matter fine-tuning than
might otherwise have been expected.

There have been several studies of the implications of prospective LHC
and/or ILC measurements for the accuracy with which the astrophysical
dark matter density could be calculated on the basis of collider
measurements\cite{Battaglia:2003ab}. These studies have emphasized
relatively favourable points in the CMSSM coannihilation region at low
$m_{1/2}$, where the dark matter fine-tuning is relatively low and the
prospective collider measurements relatively accurate. Our analysis
offers some prognosis as to which NUHM regions might be favourable for
extensions of these analyses. Clearly, the presence of NUHM points
with low dark matter fine-tuning and relatively light sparticles is
encouraging {\it a priori}.  However, we note that in general the NUHM
relic density is sensitive to the separate and independent values of
the soft Higgs masses $m_{H_{1,2}}$. Detailed consideration of their
determinations using collider data has not yet been given, as far as
we know, and we recall that the prospects for measuring directly the
masses of the heavier Higgs bosons at the LHC and ILC are limited,
though the prospects of the latter would be improved with the $\gamma
- \gamma$ option, or by going to higher energies as at CLIC.

The extension of these collider dark matter studies to the NUHM is a
large task that lies beyond the scope of this paper. However, this
exploratory study has revealed some of the prospective opportunities
and pitfalls.

\section*{Acknowledgements}

JPR would like to thank Ben Allanach for useful advice. The work of JPR
was funded under the FP6 Marie Curie contract MTKD-CT-2005-029466.
We also acknowledge partial support from the following grants:
PPARC Rolling Grant PPA/G/S/2003/00096;
EU Network MRTN-CT-2004-503369;
EU ILIAS RII3-CT-2004-506222


\begin{thebibliography}{90}


\bibitem{EENZ}
  J.~R.~Ellis, K.~Enqvist, D.~V.~Nanopoulos and F.~Zwirner,
  Mod.\ Phys.\ Lett.\  A {\bf 1} (1986) 57.
  R.~Barbieri and G.~F.~Giudice,
  Nucl.\ Phys.\ B {\bf 306} (1988) 63;

  \bibitem{hep-ph/0312378}
    D.~J.~H.~Chung, L.~L.~Everett, G.~L.~Kane, S.~F.~King, J.~D.~Lykken and L.~T.~Wang,
    Phys.\ Rept.\ {\bf 407} (2005) 1
    [arXiv:hep-ph/0312378].

  \bibitem{EHNOS}
    J. Ellis, J.S. Hagelin, D.V. Nanopoulos, K.A. Olive
    and M. Srednicki, Nucl. Phys. B {\bf 238} (1984) 453; see also
    H. Goldberg, Phys. Rev. Lett. {\bf 50} (1983) 1419.

  \bibitem{hep-ph/9506380}
    G.~Jungman, M.~Kamionkowski and K.~Griest,
    Phys.\ Rept.\ {\bf 267} (1996) 195
    [arXiv:hep-ph/9506380].


  \bibitem{etcEllis:1985jn}
    \label{etcEllis:1985jn}
    J.~R.~Ellis, K.~Enqvist, D.~V.~Nanopoulos and K.~Tamvakis,
    Phys.\ Lett.\ B {\bf 155}, 381 (1985);  M.~Drees,
    Phys.\ Lett.\ B {\bf 158}, 409 (1985).



  \bibitem{Barbieri:1987fn}
  G.~W.~Anderson and D.~J.~Castano,
  Phys.\ Rev.\ D {\bf 53}, 2403 (1996)
  [arXiv:hep-ph/9509212];
  G.~W.~Anderson and D.~J.~Castano,
  Phys.\ Rev.\ D {\bf 52}, 1693 (1995)
  [arXiv:hep-ph/9412322];
  G.~W.~Anderson and D.~J.~Castano,
  Phys.\ Lett.\ B {\bf 347}, 300 (1995)
  [arXiv:hep-ph/9409419];
    G.~G.~Ross and R.~G.~Roberts,
    Nucl.\ Phys.\ B {\bf 377} (1992) 571;
    B.~de Carlos and J.~A.~Casas,
    Phys.\ Lett.\ B {\bf 309}, 320 (1993)
    [arXiv:hep-ph/9303291];
    S.~Dimopoulos and G.~F.~Giudice,
    Phys.\ Lett.\ B {\bf 357}, 573 (1995)
    [arXiv:hep-ph/9507282];
    P.~H.~Chankowski, J.~R.~Ellis and S.~Pokorski,
    Phys.\ Lett.\ B {\bf 423}, 327 (1998)
    [arXiv:hep-ph/9712234];
    R.~Barbieri and A.~Strumia,
    Phys.\ Lett.\ B {\bf 433}, 63 (1998)
    [arXiv:hep-ph/9801353];
    P.~H.~Chankowski, J.~R.~Ellis, M.~Olechowski and S.~Pokorski,
    Nucl.\ Phys.\ B {\bf 544} (1999) 39
    [arXiv:hep-ph/9808275];
    G.~L.~Kane and S.~F.~King,
    Phys.\ Lett.\ B {\bf 451} (1999) 113
    [arXiv:hep-ph/9810374];
    J.~L.~Feng, K.~T.~Matchev and T.~Moroi,
    Phys.\ Rev.\ Lett.\  {\bf 84}, 2322 (2000)
    [arXiv:hep-ph/9908309];
    J.~L.~Feng, K.~T.~Matchev and T.~Moroi,
    Phys.\ Rev.\ D {\bf 61} (2000) 075005
    [arXiv:hep-ph/9909334];
    M.~Bastero-Gil, G.~L.~Kane and S.~F.~King,
    Phys.\ Lett.\ B {\bf 474}, 103 (2000)
    [arXiv:hep-ph/9910506];
    A.~Romanino and A.~Strumia,
    Phys.\ Lett.\ B {\bf 487}, 165 (2000)
    [arXiv:hep-ph/9912301];
    J.~A.~Casas, J.~R.~Espinosa and I.~Hidalgo,
    JHEP {\bf 0401}, 008 (2004)
    [arXiv:hep-ph/0310137];
    B.~C.~Allanach and C.~G.~Lester,
    Phys.\ Rev.\ D {\bf 73}, 015013 (2006)
    [arXiv:hep-ph/0507283];
  B.~C.~Allanach,
  Phys.\ Lett.\ B {\bf 635}, 123 (2006)
  [arXiv:hep-ph/0601089];
  P.~Athron and D.~J.~Miller,
  Phys.\ Rev.\  D {\bf 76} (2007) 075010
  [arXiv:0705.2241 [hep-ph]].






  \bibitem{Ellis:2001zk}
    J.~R.~Ellis and K.~A.~Olive,
    Phys.\ Lett.\ B {\bf 514} (2001) 114
    [arXiv:hep-ph/0105004];
    J.~R.~Ellis, K.~A.~Olive and Y.~Santoso,
    New J.\ Phys.\ {\bf 4} (2002) 32
    [arXiv:hep-ph/0202110];
    J.~R.~Ellis, S.~Heinemeyer, K.~A.~Olive and G.~Weiglein,
    JHEP {\bf 0502} (2005) 013
    [arXiv:hep-ph/0411216].



  \bibitem{hep-ph/0603095}
    S.~F.~King and J.~P.~Roberts,
    JHEP {\bf 0609} (2006) 036
    [arXiv:hep-ph/0603095];
    S.~F.~King and J.~P.~Roberts,
    Acta Phys.\ Polon.\  B {\bf 38} (2007) 607
    [arXiv:hep-ph/0609147].

  \bibitem{hep-ph/0608135}
    S.~F.~King and J.~P.~Roberts,
    JHEP {\bf 0701} (2007) 024
    [arXiv:hep-ph/0608135].

\bibitem{King:2007vh}
  S.~F.~King, J.~P.~Roberts and D.~P.~Roy,
  arXiv:0705.4219 [hep-ph].

\bibitem{Battaglia:2003ab}
    M.~Battaglia, A.~De Roeck, J.~R.~Ellis, F.~Gianotti, K.~A.~Olive and L.~Pape,
    Eur.\ Phys.\ J.\  C {\bf 33} (2004) 273
    [arXiv:hep-ph/0306219].
    E.~A.~Baltz, M.~Battaglia, M.~E.~Peskin and T.~Wizansky,
    Phys.\ Rev.\  D {\bf 74} (2006) 103521
    [arXiv:hep-ph/0602187].

  
\bibitem{oldnuhm}
  V.~Berezinsky, A.~Bottino, J.~R.~Ellis, N.~Fornengo, G.~Mignola and S.~Scopel,
  Astropart.\ Phys.\ {\bf 5} (1996) 1
  [arXiv:hep-ph/9508249];
  P.~Nath and R.~Arnowitt,
P.~Nath and R.~Arnowitt,
Phys.\ Rev.\ D {\bf 56} (1997) 2820
[arXiv:hep-ph/9701301];
  M.~Drees, M.~M.~Nojiri, D.~P.~Roy and Y.~Yamada,
Phys.\ Rev.\ D {\bf 56} (1997) 276
[Erratum-ibid.\ D {\bf 64} (1997) 039901]
[arXiv:hep-ph/9701219];
see also
M.~Drees, Y.~G.~Kim, M.~M.~Nojiri, D.~Toya, K.~Hasuko and T.~Kobayashi,
Phys.\ Rev.\ D {\bf 63} (2001) 035008
[arXiv:hep-ph/0007202];
J.~R.~Ellis, T.~Falk, G.~Ganis, K.~A.~Olive and M.~Schmitt,
Phys.\ Rev.\ D {\bf 58} (1998) 095002
[arXiv:hep-ph/9801445];
J.~R.~Ellis, T.~Falk, G.~Ganis and K.~A.~Olive,
Phys.\ Rev.\ D {\bf 62} (2000) 075010
[arXiv:hep-ph/0004169];
R.~Arnowitt, B.~Dutta and Y.~Santoso,
Nucl.\ Phys.\ B {\bf 606} (2001) 59
[arXiv:hep-ph/0102181];
V.~D.~Barger, M.~S.~Berger and P.~Ohmann,
Phys.\ Rev.\ D {\bf 49} (1994) 4908
[arXiv:hep-ph/9311269];
W.~de Boer, R.~Ehret and D.~I.~Kazakov,
Z.\ Phys.\ C {\bf 67} (1995) 647
[arXiv:hep-ph/9405342];
V.~Bertin, E.~Nezri and J.~Orloff,
JHEP {\bf 0302} (2003) 046
[arXiv:hep-ph/0210034];
H.~Baer, A.~Mustafayev, S.~Profumo, A.~Belyaev and X.~Tata,
  Phys.\ Rev.\  D {\bf 71} (2005) 095008
  [arXiv:hep-ph/0412059];
H.~Baer, A.~Mustafayev, S.~Profumo, A.~Belyaev and X.~Tata,
  JHEP {\bf 0507} (2005) 065
  [arXiv:hep-ph/0504001].
  
\bibitem{Ellis:2002wv}
J.~R.~Ellis, K.~A.~Olive and Y.~Santoso,
Phys.\ Lett.\ B {\bf 539} (2002) 107
[arXiv:hep-ph/0204192].

\bibitem{hep-ph/0210205}
  J.~R.~Ellis, T.~Falk, K.~A.~Olive and Y.~Santoso,
  Nucl.\ Phys.\  B {\bf 652} (2003) 259
  [arXiv:hep-ph/0210205].












\bibitem{Kane:1993td}
  G.~L.~Kane, C.~F.~Kolda, L.~Roszkowski and J.~D.~Wells,
  Phys.\ Rev.\  D {\bf 49} (1994) 6173
  [arXiv:hep-ph/9312272].

\bibitem{Ellis:1999mm}
J.~R.~Ellis, T.~Falk, K.~A.~Olive and M.~Srednicki,
Astropart.\ Phys.\  {\bf 13} (2000) 181
[Erratum-ibid.\  {\bf 15} (2001) 413]
[arXiv:hep-ph/9905481];
J.~Ellis, T.~Falk and K.~A.~Olive, Phys.\ Lett.\ B {\bf
444} (1998) 367 [arXiv:hep-ph/9810360];
M.~E.~G\'omez, G.~Lazarides and C.~Pallis,
Phys.\ Rev.\ D {\bf 61} (2000) 123512
[arXiv:hep-ph/9907261];
Phys.\ Lett.\ B {\bf 487} (2000) 313 [arXiv:hep-ph/0004028]
and
Nucl.\ Phys.\ B {\bf 638} (2002) 165
[arXiv:hep-ph/0203131];
T.~Nihei, L.~Roszkowski and R.~Ruiz de Austri,
JHEP {\bf 0207} (2002) 024
[arXiv:hep-ph/0206266];
S.~Mizuta and M.~Yamaguchi,
Phys.\ Lett.\ B {\bf 298} (1993) 120
[arXiv:hep-ph/9208251];
J.~Edsjo and P.~Gondolo,
Phys.\ Rev.\ D {\bf 56} (1997) 1879
[arXiv:hep-ph/9704361];
A.~Birkedal-Hansen and E.~Jeong,
arXiv:hep-ph/0210041;
H.~Baer, C.~Balazs and A.~Belyaev,
JHEP {\bf 0203}, 042 (2002)
[arXiv:hep-ph/0202076];
G.~Belanger, F.~Boudjema, A.~Pukhov and A.~Semenov,
arXiv:hep-ph/0112278;
J.~R.~Ellis, T.~Falk, G.~Ganis, K.~A.~Olive and M.~Srednicki,
Phys.\ Lett.\ B {\bf 510} (2001) 236
[arXiv:hep-ph/0102098];
J.~R.~Ellis, K.~A.~Olive and Y.~Santoso,
New Jour.\ Phys.\  {\bf 4} (2002) 32
[arXiv:hep-ph/0202110];
M.~Drees and M.~M.~Nojiri,
Phys.\ Rev.\ D {\bf 47} (1993) 376 [arXiv:hep-ph/9207234];
H.~Baer and M.~Brhlik,
Phys.\ Rev.\ D {\bf 53} (1996) 597 [arXiv:hep-ph/9508321]
and Phys.\ Rev.\ D {\bf 57} (1998) 567 [arXiv:hep-ph/9706509];
H.~Baer, M.~Brhlik, M.~A.~Diaz, J.~Ferrandis, P.~Mercadante, P.~Quintana
and X.~Tata,
Phys.\ Rev.\ D {\bf 63} (2001) 015007 [arXiv:hep-ph/0005027];
A.~B.~Lahanas, D.~V.~Nanopoulos and V.~C.~Spanos,
Mod. Phys. Lett. A {\bf 16} (2001) 1229 [arXiv:hep-ph/0009065];
J.~R.~Ellis, D.~V.~Nanopoulos and K.~A.~Olive,
Phys.\ Lett.\ B {\bf 525} (2002) 308
[arXiv:hep-ph/0109288];
J.~R.~Ellis, T.~Falk, K.~A.~Olive and M.~Schmitt,
Phys.\ Lett.\ B {\bf 388} (1996) 97
[arXiv:hep-ph/9607292];
J.~L.~Feng, K.~T.~Matchev and T.~Moroi,
Phys.\ Rev.\ Lett.\  {\bf 84} (2000) 2322
[arXiv:hep-ph/9908309];
J.~L.~Feng, K.~T.~Matchev and T.~Moroi,
Phys.\ Rev.\ D {\bf 61} (2000) 075005
[arXiv:hep-ph/9909334];
J.~L.~Feng, K.~T.~Matchev and F.~Wilczek,
Phys.\ Lett.\ B {\bf 482} (2000) 388
[arXiv:hep-ph/0004043];
    J.~L.~Feng, K.~T.~Matchev and F.~Wilczek,
    Phys.\ Lett.\ B {\bf 482} (2000) 388
    [arXiv:hep-ph/0004043];
    K.~Griest and D.~Seckel,
    Phys.\ Rev.\ D {\bf 43} (1991) 3191;
    J.~R.~Ellis, T.~Falk, K.~A.~Olive and M.~Srednicki,
    Astropart.\ Phys.\ {\bf 13} (2000) 181
    [Erratum-ibid.\ {\bf 15} (2001) 413]
    [arXiv:hep-ph/9905481].




  \bibitem{Martin:1993zk}
    S.~P.~Martin and M.~T.~Vaughn,
    Phys.\ Rev.\ D {\bf 50} (1994) 2282
    [arXiv:hep-ph/9311340].
    

  \bibitem{hep-ph/0104145}
    B.~C.~Allanach,
    Comput.\ Phys.\ Commun.\ {\bf 143} (2002) 305
    [arXiv:hep-ph/0104145].

  \bibitem{hep-ph/0112278} G.~Belanger, F.~Boudjema, A.~Pukhov and
    A.~Semenov,
    Comput.\ Phys.\ Commun.\ {\bf 149} (2002)
    103 [arXiv:hep-ph/0112278]; 
    G.~Belanger, F.~Boudjema, A.~Pukhov and A.~Semenov,
    Comput.\ Phys.\ Commun.\ {\bf 176}
    (2007) 367 [arXiv:hep-ph/0607059].  

  \bibitem{Allanach:2004rh}
    B.~C.~Allanach, A.~Djouadi, J.~L.~Kneur, W.~Porod and P.~Slavich,
    JHEP {\bf 0409} (2004) 044
    [arXiv:hep-ph/0406166].


  \bibitem{hep-ph/0703049}
    J.~P.~Miller, E.~de Rafael and B.~L.~Roberts,
    Rept.\ Prog.\ Phys.\  {\bf 70} (2007) 795
    [arXiv:hep-ph/0703049];
    G.~W.~Bennett {\it et al.} [Muon g-2 Collaboration],
    Phys.\ Rev.\ Lett.\ {\bf 92} (2004) 161802
    [arXiv:hep-ex/0401008];
G.~W.~Bennett {\it et al.}  [Muon g-2 Collaboration],
Phys.\ Rev.\ Lett.\  {\bf 89} (2002) 101804
[Erratum-ibid.\  {\bf 89} (2002) 129903]
[arXiv:hep-ex/0208001];
M.~Davier, S.~Eidelman, A.~Hocker and Z.~Zhang,
arXiv:hep-ph/0208177; see also
K.~Hagiwara, A.~D.~Martin, D.~Nomura and T.~Teubner,
arXiv:hep-ph/0209187;
F.~Jegerlehner, unpublished, as reported in
M.~Krawczyk,
arXiv:hep-ph/0208076.


  \bibitem{hfag} Heavy Flavour Averaging Group,
    www.slac.stanford.edu/xorg/hfag.

  \bibitem{hep-ex/0103042}
    K.~Abe {\it et al.} [Belle Collaboration],
    Phys.\ Lett.\ B {\bf 511} (2001) 151
    [arXiv:hep-ex/0103042];
    P.~Koppenburg {\it et al.}  [Belle Collaboration],
    Phys.\ Rev.\ Lett.\  {\bf 93} (2004) 061803
    [arXiv:hep-ex/0403004].

  \bibitem{hep-ex/0108033}
    D.~Cronin-Hennessy {\it et al.} [CLEO Collaboration],
    Phys.\ Rev.\ Lett.\ {\bf 87} (2001) 251808
    [arXiv:hep-ex/0108033].

  \bibitem{Aubert:2005cu}
    B.~Aubert {\it et al.}  [BABAR Collaboration],
    Phys.\ Rev.\  D {\bf 72} (2005) 052004
    [arXiv:hep-ex/0508004].
    B.~Aubert {\it et al.}  [BaBar Collaboration],
    Phys.\ Rev.\ Lett.\  {\bf 97} (2006) 171803
    [arXiv:hep-ex/0607071].

  \bibitem{bsgNLO}
    A.~J.~Buras and M.~Misiak,
    Acta Phys.\ Polon.\  B {\bf 33}, 2597 (2002)
    [arXiv:hep-ph/0207131];
    T.~Hurth,
    Rev.\ Mod.\ Phys.\  {\bf 75}, 1159 (2003)
    [arXiv:hep-ph/0212304].

  \bibitem{Misiak:2006zs}
    M.~Misiak {\it et al.},
    Phys.\ Rev.\ Lett.\  {\bf 98} (2007) 022002
    [arXiv:hep-ph/0609232].










  \bibitem{astro-ph/0603449}
    D.~N.~Spergel {\it et al.}  [WMAP Collaboration],
    arXiv:astro-ph/0603449.


  \bibitem{Allanach:2003jw}
    B.~C.~Allanach, S.~Kraml and W.~Porod,
    JHEP {\bf 0303}, 016 (2003)
    [arXiv:hep-ph/0302102].

  \bibitem{Belanger:2005jk}
    G.~Belanger, S.~Kraml and A.~Pukhov,
    Phys.\ Rev.\  D {\bf 72} (2005) 015003
    [arXiv:hep-ph/0502079].















































































































\end{thebibliography}
\end{document}